\documentclass{aa}  
\usepackage{graphicx}
\usepackage{array}
\usepackage{tabularx}
\usepackage{amsmath}
\usepackage{multirow}
\usepackage{natbib}
\usepackage{xcolor}
\usepackage{hyperref}
\usepackage{comment}
\usepackage[dvipsnames]{xcolor}
\usepackage[normalem]{ulem}
\usepackage[export]{adjustbox}
\hypersetup{colorlinks=true, citecolor=blue}
\usepackage[all]{hypcap}
\usepackage{orcidlink}

\usepackage{txfonts}
\usepackage{lipsum}
\usepackage{subcaption}         
\usepackage{lscape}             
\usepackage{placeins}

\newcommand{\iso}[2]{\ensuremath{{}^{#1}\mathrm{#2}}}
\newcommand{\oratio}{$^{16}$O$/^{17}$O}
\newcommand{\cratio}{$^{12}$C$/^{13}$C}

\begin{document}

   \title{Oxygen isotopic ratio in open cluster RGB stars}

   \subtitle{A new method for accurate mass and age estimations}

   \author{O. Straniero\inst{1,2}\email{oscar.straniero@inaf.it} 
        \and 
        C. Abia\inst{3}\email{cabia@ugr.es}
        \and
        I. Dom\'\i nguez\inst{3\thanks{Deceased.}}
        \and
        P. de Laverny\inst{4}
        }

   \institute{INAF-Osservatorio Astronomico d'Abruzzo, 64100 Teramo, Italy
   \and INFN-Sezione di Roma, Pz. Aldo Moro 2, 00185 Roma, Italy
   \and Dpto. F\'\i sica Te\'orica y del Cosmos. Universidad de Granada, 18071 Granada, Spain
   \and 
   Universit\'e de la C\^ote d’Azur, Observatoire de la C\^ote d’Azur, CNRS, Laboratoire Lagrange, France
   }

   \date{Received, 2026}

   \authorrunning{O. Straniero et al.}
  
   \abstract
   {Determining precise and homogeneous absolute ages of open clusters (OCs) remains a challenge in stellar astrophysics, mainly due to parameter degeneracies in photometric methods and line-of-sight extinction. Direct approaches to determine the initial masses of evolved red giant branch (RGB) stars offer a valid alternative for age estimation, an essential tool for reliably tracing the evolution of the Galactic disc. These include methods based on asteroseismology and the orbital parameters of detached binary systems, which are, however, limited to a small number of clusters.}
   {We present a novel method for estimating the ages of intermediate-age and old OCs, based on the strong initial-mass dependence of the surface oxygen isotopic ratio, \oratio, measured in RGB stars following the first dredge-up.}
   {Using updated theoretical stellar models computed with the FuNS code, we demonstrate how post-FDU CNO isotopic abundances reflect initial stellar masses. To validate this tool, we analysed archival high-resolution ($R=50\,000$) near-infrared spectra obtained with GIANO-B at the TNG for 11 RGB stars belonging to four Galactic OCs (NGC~6819, NGC~7044, NGC~7789, and NGC~6791), deriving atmospheric parameters and two isotopic ratios, specifically, \cratio\ and \oratio.}
   {For the intermediate-age clusters NGC 7789, NGC 7044, and NGC 6819, we derive ages of $1.81_{-0.30}^{+0.24}$, $1.81_{-0.26}^{+0.21}$, and $2.31_{-0.28}^{+0.26}$ Ga, respectively. For the two stars in NGC 6819, in particular, the derived initial stellar mass ($1.625_{-0.071}^{+0.063}$ M$_\odot$) and age are in excellent agreement with independent estimates from asteroseismology and eclipsing binaries. Conversely, ages estimated by means of machine learning algorithms, trained to recognise specific features in the cluster CMDs, are systematically lower. Moreover, all nine RGB stars in these clusters exhibit carbon isotopic ratio values consistent with pure post-FDU predictions, showing no evidence of extra-mixing despite being brighter than the RGB bump. Regarding the old and super-metal-rich cluster NGC 6791, the low signal-to-noise ratio and the possible line blending hinder precise mass and age constraints. With these limitations, we can only fix a lower bound of $\sim 4$ Ga for its age.}
   {The surface oxygen isotopic ratio in RGB stars provides a powerful, independent way to infer stellar ages, making high-resolution near-infrared spectroscopy of oxygen isotopes a key tool for Galactic archaeology.}


   \maketitle
\nolinenumbers

\section{Introduction}\label{sect:intro}\label{sect:1}
\nolinenumbers
Open clusters (OCs) are dynamically loose stellar systems formed from the same giant molecular cloud. Because their constituent stars share similar distance, initial chemical composition, and age, OCs serve as ideal ``laboratories'' for testing stellar evolution theories and probing the properties of the Galactic disc. Scattered throughout the Milky Way's disc, these systems act as excellent tracers of its chemical and dynamical history. They allow us to map the Galactic metallicity gradient, reconstruct the star formation history, and understand the mechanisms driving radial migration and disc heating over time \citep[e.g.,][]{fri95, dia02, net16, pra23}.

OCs cover a wide age range, from a few tens of Ma to a few Ga. However, the use of OCs to accurately reconstruct the timeline of the Galactic disc relies heavily on our ability to derive precise and homogeneous age estimates. Despite their importance, determining the absolute ages of OCs remains a challenging task in modern astrophysics, as different methodologies are often hampered by distinct observational and physical limitations.

Historically, the most widely used technique for estimating cluster ages is isochrone fitting on Colour-Magnitude Diagrams (CMDs). This method relies on comparing some age dependent features of the CMD, specifically the main sequence turn-off (MSTO), with theoretical isochrones \cite[e.g.,][]{str91, dom99, bre12,spa17}.
This method is conceptually straightforward and can be applied to large photometric datasets spanning broad populations of stars. Nevertheless, isochrone fitting is highly sensitive to degeneracies between age, metallicity, distance, and interstellar extinction. In the Galactic plane, where OCs are primarily located, high and differential reddening can severely distort CMDs. Furthermore, OC distances recently derived from  parallaxes available after Gaia DR3 exhibit large star-to-star scatter, particularly for objects placed beyond $\sim 2.5$ kpc from the Solar System, making their application for age determination difficult \citep[e.g.][]{bai21}. Contamination of stellar samples by field stars and the presence of a high fraction of unresolved binaries pose additional challenges. All of these uncertainties directly translate into significant age discrepancies. The use of machine learning techniques allows for simultaneous and therefore more consistent fits of distance, reddening, and age; however, it cannot break parameter degeneracies or eliminate photometric and other uncertainties \citep[e.g.,][]{cav24,hunt24,chan20,sod14}.

In recent years, asteroseismology has emerged as a powerful alternative to determine stellar ages. In general, this technique uses space-based photometry to detect the frequencies of stellar pulsations \cite[e.g., Kepler, K2, TESS;][]{cha13}. For solar-like stars, main sequence stars with a radiative core, the difference among the frequencies of low angular degree $p$ modes (l=0, 1, 2) depends on the core composition. In practice, stellar  ages can be determined by measuring the H consumption. For younger stars, those with an H-burning convective core, mixed modes can be used. This technique has been applied to a few open clusters \citep{bro25}. {Incidentally, correlating hydrogen consumption with the age of the star strictly requires the use of stellar models, whose uncertainties, especially those related to the treatment of convection and core overshooting, are reflected in the final result.}

Alternative methods consist of measuring the masses of main-sequence stars in the turn-off region or those of more evolved stars, e.g., RGB stars, by leveraging an age-versus-mass relationship. In the case of binaries, current masses can be derived from orbital parameters, such as period, radial velocities, and inclination. Ideal systems are eclipsing binaries (EBs), for which inclination is not an issue \citep{bre16,bro21,yak26}.
Asteroseismology can also provide precise masses and radii of RGB stars in galactic OCs \cite[e.g.,][]{brogaard_2023A&A,tay25,will26, bas11, Reyes_2025MNRAS, Palakkatharappil_2023A&A, how26}.
Potentially, these methods provide highly precise physical properties for individual cluster members, yielding some of the most precise absolute ages to date. However, these methods only yield the present-day mass of these stars, while their age is determined by their initial main-sequence mass. Consequently, the unknown history of mass-loss remains a major limit. Moreover, not many clusters present RGB stars in EBs, while asteroseismology is observationally expensive, requiring long-term, high-cadence time-series photometry, restricting its application to a relatively small sample of bright, nearby clusters.  

The limitations of these standard techniques create a significant bottleneck, especially in the era of large-scale Galactic surveys. Optical photometry often fails to penetrate the heavily obscured, dust-enshrouded regions of the inner Galactic disc, where a vast population of OCs resides. Although \textit{Gaia} astrometry has revolutionised cluster member identification \citep[e.g.,][]{chan20,hunt24,cav24}, deriving accurate ages remains a challenge.

To overcome these biases and build a more complete picture of the Galactic disc, alternative diagnostics are required. Near-Infrared (NIR) spectroscopy offers a powerful solution to this problem for old and intermediate age clusters. NIR wavelengths are significantly less affected by interstellar extinction ($A_{\text{NIR}} \approx 0.1 \times A_V$), allowing us to probe deeper and further and further into the Galactic plane. Furthermore, high-resolution NIR spectra provide a wealth of chemical and atmospheric information from stellar surfaces that can bypass traditional photometric degeneracies.

In this paper, we present a novel method for estimating OC ages ranging between 1 and 10 Ga by using the strong dependence on the initial stellar mass of the oxygen isotope ratio, \oratio, measured in the atmospheres of RGB stars. We will show how this approach can provide robust and homogeneous age constraints for clusters across a wide range of Galactic environments.

Section 2 discusses the relationship between the \oratio~and RGB star masses, introducing our method for determining the age. The sample and chemical analysis of the OC RGB stars employed to demonstrate the robustness of the method is presented in Section 3, while Section 4 is dedicated to the discussion of our results. The conclusions follow.

\section{RGB stars: models and predictions}\label{sect:2}

\begin{figure}.  
   \centering
   \includegraphics[width=8 cm]{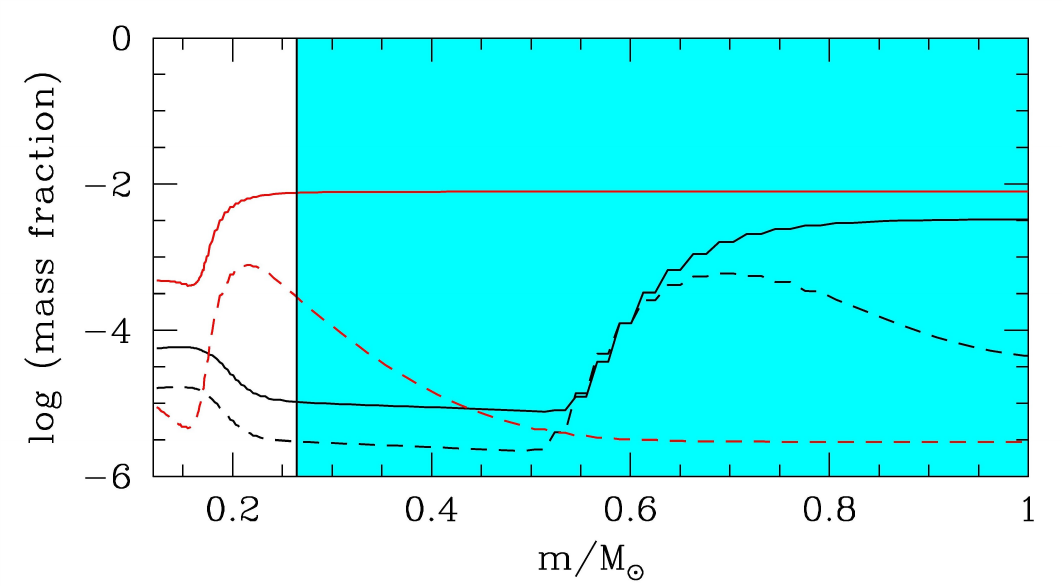}
   \caption{Abundance profiles of carbon and oxygen isotopes in a 1.6 M$_\odot$ model with solar composition, just before the FDU. The figure shows the innermost region of the H-rich envelope. Black lines refer to \iso{12}{C} (solid) and \iso{13}{C} (dashed), while red lines represent \iso{16}{O} (solid) and \iso{17}{O} 
   (dashed). Later on, the convective envelope extends inward, eventually reaching the layers previously processed by nuclear burning. The coloured area indicates the maximum penetration of the outer convection.
   } 
   \label{fig:fig:1dup}
\end{figure}

After the central depletion of hydrogen, low-mass stars move toward the giant branch; the He-rich core contracts, and a hydrogen-burning shell controlled by the CNO cycle forms, while the H-rich envelope expands and cools. Under these conditions, the convective instability, which during the core-H burning phase was confined to the outermost and coldest region of the star, can extend inward until it reaches layers where the chemical composition has been altered by nuclear burning. This phenomenon, known as the first dredge-up (FDU), is illustrated in Figure \ref{fig:fig:1dup} for the case of a $1.6 \text{ M}_\odot$ star with a solar chemical composition. This figure shows the stratification of carbon and oxygen isotopes in the interior of the star, just before the external convection penetrates to reach the layers where the production and destruction of \iso{12}{C} and \iso{13}{C} are in equilibrium, \iso{17}{O} is produced, while $^{16}$O is only marginally consumed. As a result, when the star begins to ascend the giant branch, its surface chemical composition appears altered. In particular, the $^{12}\text{C}/^{13}\text{C}$ ratio is reduced from about 90 to approximately 25, while the $^{16}\text{O}/^{17}\text{O}$ ratio drops from 2700 to approximately 500\footnote{Solar system $^{12}$C/$^{13}$C and $^{16}$O/$^{17}$O isotopic ratios are $89$ and $2696$, respectively \citep{lod09}.}. 

The most relevant aspect is that for stars with masses between 1 and 2 $\text{M}_\odot$, while the changes in the carbon isotopic ratio are relatively small, the oxygen ratio is particularly sensitive to variations in the initial mass or, specifically, the mass of the zero-age main sequence (ZAMS). {To understand this difference, one must look at the maximum inward penetration of the convective envelope during the FDU. As clearly shown in Fig.~\ref{fig:fig:1dup}, convective mixing extends deep into the region where $^{12}\text{C}$ is depleted and $^{13}\text{C}$ is produced. Note that the  $^{13}\text{C}$ peak occurs at $m/M_\odot \approx 0.7$, while convection extends down to $m/M_\odot \approx 0.26$, reaching the zone where the abundances of the two C isotopes are set by the complete equilibrium of the CN cycle. Although Fig.~\ref{fig:fig:1dup} refers to the M $= 1.6$ M$_\odot$ case, this feature is common to all models between 1 M$_\odot$ and 2 M$_\odot$. In contrast, the convective envelope does not reach the region where $^{16}\text{O}$ begins to be consumed, but merely grazes the outer tail of the $^{17}\text{O}$ production peak. In this case, the maximum $^{17}\text{O}$ abundance lies below the maximum penetration depth of the convection, precisely at $m/M_\odot \approx 0.22$. Thus, the position of the $^{17}\text{O}$ peak relative to the maximum depth reached by the convective envelope varies with stellar mass. This occurrence explains the strong sensitivity of the surface $^{17}\text{O}$ abundance to the initial mass.}
It also makes the measurement of the \oratio~  isotopic ratio in RGB stars a potential estimator of their initial mass and, in turn, of their age. 

\begin{figure}
   \centering
   \includegraphics[width=8 cm]{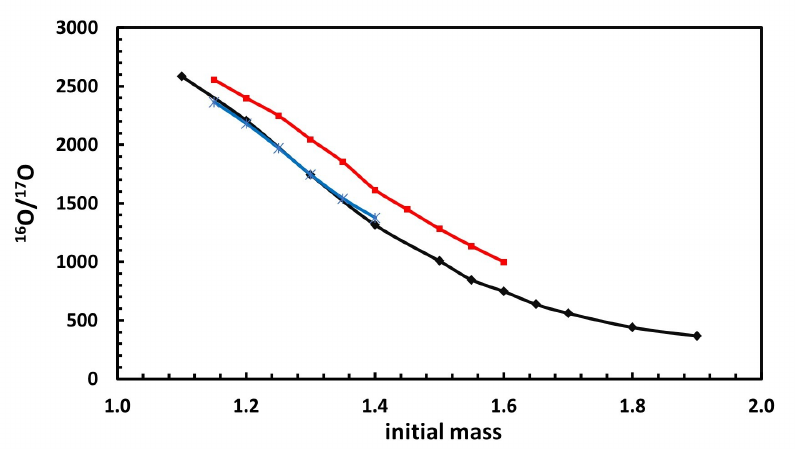}
   \includegraphics[width=8 cm]{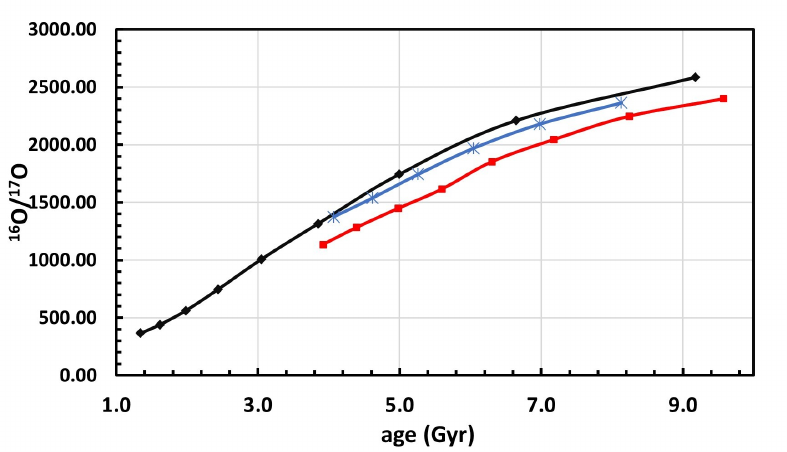}
   \caption{\oratio~ after the FDU as a function of the ZAMS mass (in M$_\odot$) and corresponding age. Black and red lines refer to models with solar He abundance ($Y=0.275$), $\text{[Fe/H]}=0$ and $\text{[Fe/H]}=0.3$, respectively, while the blue line refers to model with He enhanced ($Y=0.32$) and $\text{[Fe/H]}=0.3$. The initial chemical composition is solar scaled.
   } 
   \label{fig:age_mass}
\end{figure}

The predicted post-FDU \oratio~ for stars with mass between 1 and 2 M$_\odot$ and [Fe/H]$=0$ and $+0.3$ is shown in the upper panel of Fig. \ref{fig:age_mass}, while the lower panel reports the corresponding \oratio-age relations.   

\begin{figure}
   \centering
   \includegraphics[width=8 cm]{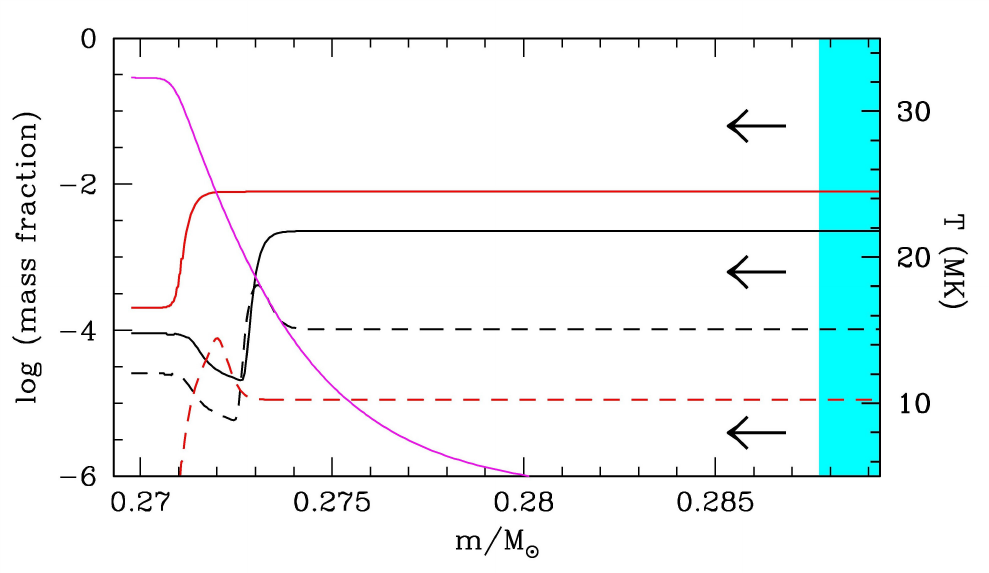}
   \caption{Abundance profiles of carbon and oxygen isotopes in a 1.6 M$_\odot$ model with solar composition, just after the RGB bump ($\log L/L_\odot\sim 1.85$). The figure shows the innermost region of the H-rich envelope. Black lines refer to \iso{12}{C} (solid) and \iso{13}{C} (dashed), while red lines represent \iso{16}{O} (solid) and \iso{17}{O} (dashed). The magenta line represents the temperature profile, whose values are reported on the right vertical axis. The coloured area indicates the convective envelope, while the arrows point to the region where an extra-mixing process might be at work. 
   } 
   \label{fig:mod4900}
\end{figure}

A potential issue for this mass/age estimator is the possible activation of non-convective mixing processes during the RGB evolution, which may modify the surface abundance of oxygen isotopes. The predicted \cratio~ is generally in agreement with that measured in post-FDU stars, but an extra-mixing process may occur on the RGB, further lowering the surface ratios in stars with masses smaller than 2 M$_\odot$ and brighter than the characteristic bump of the RGB luminosity function \citep[e.g.,][]{sweigart79,dearborn76,charbonnel1994A&A}. Indeed, stars in bright RGB often exhibit particularly low carbon isotopic ratios, specifically $\text{\cratio} < 15$ \citep[e.g.][]{brow89,smi02,pil03,sm07,abia2012,maa19,agu23}. This finding provides evidence that an extra-mixing process may be active during the RGB. However, the FDU leaves a steep molecular-weight gradient that prevents mixing from propagating inward to the layers where nuclear reactions alter the CNO composition. Eventually, this barrier is erased once the H-burning shell reaches the FDU discontinuity, allowing extra-mixing to penetrate further inward. Various possible engines for this additional mixing have been invoked, among them, differential rotation, gravity waves, thermohaline circulations, and magnetic buoyancy \citep[e.g.][]{eggleton06,den10,pal111,lai11,joh12,har19,tay22,blo26,oca26}.
Fig.~\ref{fig:mod4900} shows the C and O isotopic profiles in the region below the convective envelope for a model of 1.6 $M_\odot$ (solar composition) just after the removal of the molecular-weight barrier. Note that the temperature at the inner border of the convective envelope is well below the threshold for the activation of the CNO cycle. Carbon isotopes are modified only if the extra-mixing process reaches layers where $T > 15$~MK, while a change in oxygen abundance requires higher temperatures, specifically $T > 20$ and $22$ MK for \iso{17}{O} and \iso{16}{O}, respectively. It should be noted that a mixing deep enough to alter \oratio~ would also supply fresh H fuel to the innermost layers of the burning shell, likely affecting the RGB luminosity function.    
 
All models we used for this work were specifically calculated using the FuNS (Full Network Stellar evolution) code (see \cite{straniero_2020} for details). FuNS is a 1D hydrostatic code widely used for computing stellar models of any mass and complete evolutionary sequences up to the final stages. Probably the most relevant feature for the calculations discussed here is the possibility of simultaneously solving the stellar structure equations and those governing the variations in the internal chemical composition, the latter being due to both the activation of nuclear processes and mixing induced by convection or other instabilities. A fully coupled solver is particularly important when mixing extends down to a region where nuclear reactions are active. This is the case of extra-mixing possibly active in bright RGB stars. A full nuclear network of 27 isotopes is adopted, specifically: $^1$H, $^2$H, $^3$He, $^4$He, $^6$Li, $^7$Li, $^7$Be, $^{12}$C, $^{13}$C, $^{14}$N, $^{15}$N, $^{16}$O, $^{17}$O, $^{18}$O, $^{19}$F, $^{20}$Ne, $^{21}$Ne, $^{22}$Ne, $^{23}$Na, $^{24}$Mg, $^{25}$Mg, $^{26}$Mg, $^{26}$Al, $^{27}$Al, $^{28}$Si, $^{29}$Si and $^{30}$Si. Fusion reaction rates are from the STARLIB compilation \citep{sal13}, except for a few key reactions of the CNO cycle. Specifically, based on recent studies of the R-matrix that include new low-energy cross-section measurements \citep{Skowronski_2023PhRvL, rapagnani_2025PhRvC}, we have updated the rates for the following reactions: $^{12}$C$(p,\gamma)^{13}$N , $^{13}$C$(p,\gamma)^{14}$N, $^{17}$O$(p,\gamma)^{18}$F and $^{17}$O$(p,\alpha)^{14}$N. {The reaction responsible for $^{17}\text{O}$ production, i.e., $^{16}\text{O}(p,\gamma)^{17}\text{F}$, has also been updated following the latest underground measurements obtained by the LUNA collaboration \citep{Roob_luna2026}. These papers showed that recent experimental measurements of key nuclear cross sections in the CNO cycle have minimized uncertainties, making their impact on stellar-model predictions for C and O isotopic ratios negligible.}

In the following section, we illustrate our proposed method to derive the initial mass ($M_{ini}$) and, in turn, the age of a typical RGB star, namely $\alpha$~Boo (Arcturus), based on the simultaneous fit of both \cratio~ and \oratio~ ratios. We chose this star not so much because it is the brightest red giant ($V \sim 0.5$), but rather because its low measured $^{12}\text{C}/^{13}\text{C}$ ratio, combined with a distinctly non-solar chemical composition (in particular $[\text{O}/\text{Fe}] \sim 0.4$), makes Arcturus the most challenging case to analyse with the method proposed here. For this reason, it serves as the most significant benchmark to demonstrate the potential of this method.

\subsection{Initial mass and age of Arcturus}\label{sec:arcturus}

{
The primary goal of this paper is to present a method for determining masses and ages of RGB stars through the measurement of the \oratio~ ratio. In this context, Arcturus, a bright RGB star in the Bootes constellation, represents an excellent benchmark. Indeed, it encompasses all the potential challenges that may be encountered when applying our method. For instance, it is quite metal-poor, its oxygen abundance is significantly enhanced relative to iron, and its carbon isotopic ratio is substantially lower than the value expected after the FDU, which raises the issue of extra-mixing.

In general, CNO isotopic ratios, particularly the \cratio, are good tracers of the interplay between nuclear burning and mixing processes, such as convection or other dynamical and secular instabilities. Therefore,
simultaneous measurements of carbon and oxygen isotopic can be used to determine the extent of the mixing or, equivalently, the maximum temperature to which the material observed at the stellar surface has been exposed, while simultaneously providing constraints on the initial mass of the star. When applied to nearby RGB stars, which are well characterized by means of photometric and spectroscopic studies, this procedure may provide further insight into the extra-mixing process.
} 

\begin{figure}
    \centering
    \includegraphics[width=8 cm]{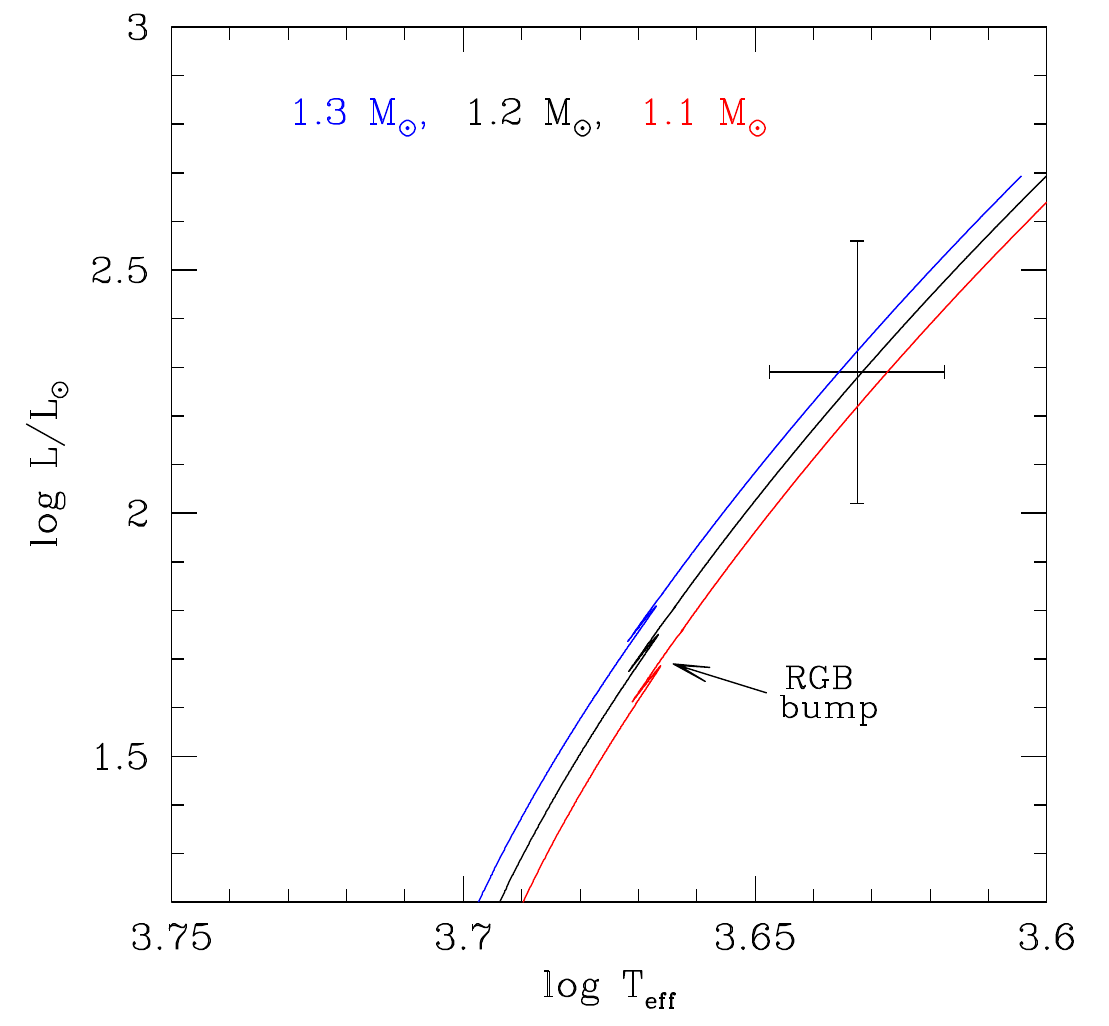}
    \caption{RGB evolutionary tracks for M=1.1, 1.2 and 1.3 M$_\odot$ and solar composition are compared to the measured luminosity-effective temperature of $\alpha$ Boo.}
     \label{fig:hr}
\end{figure}

Arcturus ($\alpha$ Boo is the brightest star in the northern celestial hemisphere (apparent visual magnitude $m_V=-0.05$ mag, $196\pm 21$ L$_\odot$, T$_{\rm {eff}}=4290\pm 50$ K, log g$=1.5\pm 0.1$, and distance 11.26 pc).
Current estimates of its mass are mostly based on photometric observations and range between $1$ and $2$ $\text{M}_\odot$ (see, e.g., \cite{abia2012} and references therein). Fig. \ref{fig:hr} shows the location of $\alpha$ Boo in the $\log L-\log T_{\text{eff}}$ plane, together with the theoretical evolutionary paths of the $M=1.1$, $1.2$, and $1.3$ $\text{M}_\odot$ models. Note the poor mass constraint obtained from this plot. Moreover, this star is substantially brighter than the RGB bump. 

\begin{table}
    \centering
     \caption{Main spectroscopic abundances  and isotopic ratios of $\alpha$-Boo from \cite{abia2012}.}
    \begin{tabular}{l c}
    \hline\hline
     Specie       &  Abundance/isotopic ratio   \\
    \hline
    Fe & $6.98\pm 0.07$             \\
    C  & $8.06\pm 0.09$    \\
    N  &  $7.67\pm 0.16$     \\
    O  &  $8.76\pm 0.17$    \\
    $[\text{Fe}/\text{H}]$    & $-0.50\pm 0.07$        \\
    $[\text{C}+\text{N}/\text{Fe}]$  & $0.00\pm 0.12$      \\
    $[\text{O}/\text{Fe}]$    & $0.42\pm 0.15$     \\
    \cratio   & $9\pm 2$   \\
    \oratio   &  $3030\pm 530$   \\
    \hline\hline
    \end{tabular}
    \label{tab:arcturus}
\tablefoot{The absolute abundances in the table are given as A(X)$=$log N(X)/N(H)$ + 12$, where N(X) is the abundance by number of the element X in the scale  log N(H)$\equiv 12$. As usual, [X/Y]$= \rm{log (N(X)/N(Y))_\star - log (N(X)/N(Y))_\odot.}$}
\end{table}

The present day composition, as deduced from optical and IR spectra, is summarised in Tab. \ref{tab:arcturus} from \cite{abia2012} (see also \cite{fan21}).
Overall, Arcturus presents a sub-solar metallicity, with oxygen enhanced and solar-scaled C+N. Typically, stars with lower-than-solar metallicity originate from a gas in which O is overabundant with respect to iron, while C and N scale with iron. During RGB evolution, the abundances (by number) of C and O might decrease while that of N increases, but their sum is conserved. This is a well-known property of the CN+NO cycles. In this framework, the most straightforward interpretation of the observed abundances is that oxygen reflects the primordial composition, whereas the observed C depletion, coupled with the nitrogen enhancement, is the consequence of mixing with material processed by the CN cycle only. The evidence of a substantially solar-scaled C+N abundance confirms that the oxygen depletion was marginal, if any. 

According to this analysis, our stellar model computations start with an O-enhanced composition: 
$\text{[Fe/H]}=-0.50$ and $\text{[O/Fe]}=+0.4$. The overabundance of oxygen with respect to iron is essentially driven by the primary and most abundant isotope \iso{16}{O}. This specy is preferentially produced in the He and C-burning regions of massive stars, the progenitors of core-collapse supernovae, through the chain $3\alpha\rightarrow{\rm ^{12}C}+\alpha\rightarrow{\rm^{16}O}$. In contrast, \iso{17}{O} is synthesised in the H-burning shells of stars across a wider mass range; for this reason, its abundance scales as a secondary element with iron \citep[see e.g.][]{koba11}. Hence, the initial $^{16}\text{O}/^{17}\text{O}$ isotopic ratio depends exclusively on the overabundance of the primary isotope, in our case, adopting the value [O/Fe] in Table 2:

\begin{equation}
    \left(\frac{^{16}\text{O}}{^{17}\text{O}}\right)_{\star} = \left(\frac{^{16}\text{O}}{^{17}\text{O}}\right)_{\odot} \times 10^{[\text{O}/\text{Fe}]} \approx \left(\frac{^{16}\text{O}}{^{17}\text{O}}\right)_{\odot} \times 2.6 \approx 7090
\end{equation}
\noindent
where the solar \oratio~ is 2696 \citep{lod09}.
Regarding the two stable carbon isotopes, both show  a primary-like scaling with metallicity and, since [C/Fe]$\approx 0.0$, our models start with a solar $^{12}\text{C}/^{13}\text{C}$ (89).   

 As shown in Fig. \ref{fig:hr}, the observed luminosity and effective temperature of Arcturus provide a first rough estimation of its initial mass: $\sim 1.2$ M$_\odot$. 
Following the FDU, \oratio\ decreases to 2320, 3700, and 5730 for models 1.3, 1.2, and 1.1 M$_\odot$, respectively. Note how the post-FDU value strongly depends on the initial mass. These variations in the oxygen ratio are due to the internal production and dredge-up of \iso{17}{O}, while \iso{16}{O} remains practically unchanged. In contrast, \cratio\ decreases to 27, 28.5 and 29.5 for the models of 1.3, 1.2 and 1.1 M$_\odot$, respectively. In this case, the influence of the mass is rather weak, but all these values are significantly higher than the \cratio\ $\sim 9$ found in Arcturus. This discrepancy is commonly considered evidence that an extra-mixing process is active below the convective envelope of stars in the brightest portion of the RGB ($\log L/L_\odot \geq 1.8$).

In order to simulate this  process, we extend the mixing below the convective envelope, down to a temperature $T_{max}$. In the zone where extra-mixing occurs, the average mixing velocity ($v_{mix}$) is assumed to be constant and lower than that of the deeper convective-envelope layer. The extra-mixing process is then switched on just after the occurrence of the RGB bump.
In practice, this simple model depends on three free parameters: $f = v_{mix}/v_{CE}$, which is the ratio between the mixing velocity within the extra-mixed zone and the average turbulent velocity at the inner border of the convective envelope\footnote{As computed according to the mixing-length theory.}, $T_{max}$ and the initial mass ($M_i$). Actually, the resulting \cratio\ and \oratio\ depend weakly on $f$ (if $f > 10^{-4}$), while they are much more sensitive to a change in $T_{max}$ and $M_i$. In particular, we found that the measured \cratio~ provides a strict constraint to $T_{max}$, while \oratio~ is more sensitive to $M_i$. 
Fitting the observed \cratio\ and \oratio\ yields $T_{\text{max}} = 20.03 \pm 0.02$\,MK and $M_i = 1.23 \pm 0.02$ M$_\odot$. 
{The fitting procedure is detailed in Appendix~\ref{A4}. The quoted uncertainties represent the propagation of the \cratio\ and \oratio\ errors, whereas uncertainties in the initial chemical composition (specifically, [O/Fe] and [Fe/H]) were not considered. Naturally, varying these two additional parameters would require computing a significantly larger grid of models. We estimate that the overall uncertainty on the initial mass may be approximately twice that obtained solely from variations in the isotopic ratios.
}

Some interesting insights can be drawn from this analysis.
In the case of Arcturus, the maximum temperature attained by the extra-mixing process is high enough to slightly increase the surface abundance of \iso{17}{O}, while \iso{16}{O} remains unaffected (see also the previous comment on Fig. \ref{fig:mod4900}). We reiterate that $T_{max}$ is primarily constrained by the measured \cratio. In practice, when  \cratio\ is larger than $\sim 11$, the required $T_{max}$ is less than $20$ MK and the \oratio\ does not change. As shown in the next section, the \cratio\ values measured in the RGB stars in our sample are generally large enough to imply moderate extra-mixing, if any. In this case, the measured \oratio\ is solely a consequence of the FDU and, in turn, depends on just one free parameter: the initial mass. Nonetheless, the complete procedure described here for Arcturus can also be applied in the future to other cases where the effects of extra-mixing cannot be neglected.

\section{Spectroscopic data, abundance determination and luminosity of open cluster RGB stars}\label{sect:3}
Data were acquired on the Italian 3.6 m TNG located at the
Observatorio del Roque de los Muchachos on La Palma (Canary
Islands, Spain), using the GIANO-B (R$ = 50 000$, $\lambda\lambda= 0.97 -
2.5$ $\mu$m) spectrograph. The spectra were taken from the TNG archive\footnote{http://archives.ia2.inaf.it/tng/}, where they
are publicly available. The spectra were already reduced, including bad pixel removal, 
sky and dark subtraction, flat-fielding, optimal spectrum extraction, and wavelength
calibration. Stellar spectra were also corrected for telluric contamination using a
standard early-type star observed during the same night. The S/N ratio, achieved in the
observed spectral region, ranges from 150 to more than 400. The observations were done for other purposes,
namely the study of the fluorine abundance in young open cluster stars as a function of their
galactocentric distance. For more details on observations, data acquisition, and reduction, we refer to the original article by \cite{bij24}.

The selected stellar sample comprises 11 RGB stars in four OCs: NGC 6791, NGC 6819, NGC 7044, and NGC 7789. NGC 6791 represents the older component of our sample, whereas the other three are intermediate-age clusters. However, the age estimations differ considerably in the literature. Table 3 summarises the most relevant age determinations for these OCs over the last 25 years, together with the method used to derive them. The dispersions of the available age estimates are also reported. In the overwhelming majority of studies, the classical isochrone fitting method was used. Discrepancies in the derived ages mainly arise from differences in the stellar evolutionary codes used to compute the isochrones, the adopted metallicity [Fe/H], extinction $A_V$ (or reddening $E(B-V)$), distance, and the identification of the main-sequence turn-off region, which in these clusters is often poorly populated and heavily dispersed, largely due to a high fraction of unresolved binaries. 
Most recent analyses use artificial neural network (ANN) or machine learning (ML) codes trained on synthetic colour-magnitude diagrams. These techniques aim to simultaneously estimate all cluster parameters, including their ages (see, e.g. \cite{cant20,hunt24,cav24}). The substantial differences in the ages estimated by the authors using this type of code cannot be ignored, even when the same set of isochrones was used for training. 
When available, mass and radii measurements for binary components can also provide fairly accurate ages. For two clusters, NGC 6819 and NGC 6791, asteroseismic data from RGB stars were also used to constrain their current masses and, in turn, their ages. Overall, the data presented in Table 3 clearly show that age estimates for Galactic open clusters remain rather uncertain with age dispersions from 25 to almost 100$\%$ from different studies.

\begin{table}
\caption{Open cluster ages from recent studies.}             
\tiny
\centering          
\begin{tabular}{c c c }     
\hline\hline       
                      
 Age (Ga) & Method & Reference \\ 
\hline 
       \multicolumn{3}{c}{NGC 6791} \\  
         8.20 & B + Iso(1) & \cite{gru08} \\
         7.70 & B + Iso(2) & \cite{gru08}\\
         9.00 & B  + Iso(3) & \cite{gru08}\\
         8.00 & WDC & \cite{gar10}\\
         8.30 & B+Iso(1) & \cite{bro21,bro12}\\
         9.80& ML+Iso(4) & \cite{cant20}\\
         9.54& ML+Iso(4) & \cite{cav24}\\
         9.57& ML+Iso(4) & \cite{hunt24}\\
         8.20& Iso(4)& \cite{amh25}\\
         8.46& Iso(5) & \cite{duf26}\\
         7.00&  AST+Iso(2)  & \cite{bas11}\\
         7.23 & Iso(4)   & \cite{dia21} \\
         $\delta = 2.80$ (33\%) &    & \\
\hline             
        \multicolumn{3}{c}{NGC 6819} \\  
          1.66 & ML+Iso(4) & \cite{cant20}\\
          1.99 & ML+Iso(4) & \cite{cav24}\\
          1.65 & ML+Iso(4) & \cite{hunt24}\\
          2.38 & B+Iso(1,2,4)& \cite{bre16}\\
          2.0-2.4 &  AST+Iso(2) & \cite{bas11}\\
          2.63 & Iso(4)   & \cite{dia21} \\
          $\delta = 0.98$ (46\%) & & \\
\hline           
          \multicolumn{3}{c}{NGC 7044} \\
           0.67 & ML+Iso(4) & \cite{cant20}\\
           1.34 & ML+Iso(4) & \cite{cav24}\\
          0.66 & ML+Iso(4) & \cite{hunt24}\\
          1.04 &  Iso(7)  & \cite{sal04}\\
          1.32 & ML+Iso(4) &  \cite{dia21}\\
          1.90 &  ML+Iso(4)  &  \cite{kha13}\\
    
         1.33  & Iso(4)   & \cite{dia21} \\
         $\delta = 1.24$ (97\%) & & \\
\hline        
          \multicolumn{3}{c}{NGC 7789} \\
          1.54& ML+Iso(4)& \cite{cant20}\\
          1.33& ML+Iso(4)& \cite{cav24}\\
         1.54& ML+Iso(4)& \cite{hunt24}\\
          1.26 & B   & \cite{yak26}\\
          1.62 & Iso(4)   & \cite{dia21} \\   
          $\delta = 0.36$ (25\%) & \\
\hline\hline                  
\end{tabular} 
\tablefoot{The last row in each cluster indicates the corresponding age dispersion from the literature. B: binary masses; Iso: isochrone fitting: (1) Victoria-Regina, (2) Yale, (3) BaSTi, (4) PARSEC, (5) \cite{dot08}, (6) \cite{ber94}, (7) \cite{sal98}; WDC: White dwarf cooling; ML: Machine learning. AST: Asterosismology.}
\label{tab:OCage}
\end{table}

\subsection{Chemical analysis}\label{sec:chemical}
Initially, we used the same stellar parameters
(T$\rm _{eff}$, gravity, average metallicity [M/H], microturbulence and CNO abundances)
previously derived in the study by \cite{bij24}. 
Of particular relevance is the determination
of the CNO abundances. Therefore, we checked the abundance values originally adopted for these elements. For this we compute synthetic spectra in LTE with the Turbospectrum code v19.5 \citep{alv98}\footnote{https://github.com/bertrandplez/Turbospectrum2019.} using a
model atmosphere for each star interpolated in the MARCS grid of models \citep{gus08} with the
corresponding stellar parameters. Synthetic spectra were convolved with a Gaussian function, which includes the instrumental profile and the macroturbulence parameter. Typically, this ranges from 6 to 8 kms$^{-1}$. Firstly, the abundance of oxygen was derived from some OH lines present in the $\lambda\lambda\sim 1.55-1.60$ $\mu$m region. Then, with this oxygen abundance we estimate the carbon abundance from many of the CO lines also present in this region of the spectrum. Finally, the nitrogen abundance was derived (keeping the C and O abundances fixed) from a few weak CN lines. The molecular and atomic line list used in this region is that of the APOGEE project \citep{maj17,smi21}. Since in cool stars, when the C/O ratio is lower than unity, the available carbon in the atmosphere is mostly locked into CO, 
this procedure was repeated iteratively until the combination of CNO abundances derived gave good fits (using the $\chi^{2}$ method of minimisation) to the selected molecular features OH, CO, and CN. We tried to fit not only these molecular lines, but also the global observed spectrum of the star in the region using the $\chi^{2}$ method. Figure 5 shows an example of this, where some metallic lines used to derive the average metallicity are shown. In case of significant disagreement between the observed and theoretical spectra, we modified the stellar parameters initially adopted until a good fit was found.

Once preliminary values for CNO abundances and average metallicity from some FeI, SiI, NiI and TiI lines have been determined in the region 1.55-1.60 $\mu$m, we study the spectral region $\lambda\lambda\sim 2.33-2.35$ $\mu$m where the $^{12}$C$^{17}$O lines are found for the derivation of the $^{16}$O/$^{17}$O ratio. This spectral region is dominated by strong CO band absorptions; nevertheless, there are also a few weak CN and C$_2$ lines, and some NaI, ScI, and other weak metallic lines, as well as some HF lines. Synthetic spectra were computed  to derive again C and O abundances from a few weak, unblended C$_2$, CN and CO lines. For most of the stars, the CN lines used in this spectral range are almost insensitive to middle variations in the nitrogen abundance. The atomic and molecular line lists used  are the same as in \cite{hed13, abi15} and \cite{abi7}. We refer to these studies for details. Careful selection of $^{12}$C$^{17}$O lines was performed in the region 2.3 $\mu$m; in particular, we selected only weak lines not significantly
affected by blends and sensitive to variations in the abundance of $^{17}$O. The list of selected lines can be found in Table 2 of \cite{abi7}. These lines are only marginally affected by the actual $^{12}$C/$^{13}$C ratio and nothing at all by the $^{16}$O/$^{18}$O one, provide this latter ratio is higher than 500\footnote{Note that the solar $^{16}$O/$^{18}$O ratio is 490 \citep{lod09} and that the expected post-FDU value is larger than solar for stars in the mass range $1.2-4.0$ M$_\odot$ with solar composition. The existence of any extra-mixing during the RGB would deplete $^{18}$O, increasing this ratio.}. Thus, despite that there are some useful $^{12}$C$^{18}$O lines in this spectral region, we do not derive the $^{16}$O/$^{18}$O ratio. We checked whether the C and O abundances derived in this region are consistent with those obtained from the analysis of the $\lambda\lambda\sim 1.55-1.60$ $\mu$m region. In case of a larger discrepancy than the uncertainties (see below), we restart the procedure modifying the stellar parameters slightly (mainly T$_{\rm{eff}}$ and log g) in an iterative process until convergence is achieved from both spectral regions. We note that the OH lines and CO-band lines used in the $1.55-1.60$ $\mu$m region are particularly sensitive to changes in effective temperature and gravity, respectively. In general, we found good agreement between the CNO abundances derived from both spectral ranges, and in most stars only a few iterations were needed. The final stellar parameters and CNO abundances derived in this way are shown in Table 3. Good agreement was found between these values and the corresponding figures determined in \cite{bij24,bij25}, although the microturbulence velocities determined here are systematically larger than those derived by these authors.

\begin{figure}
   \centering
   \includegraphics[width=8 cm]{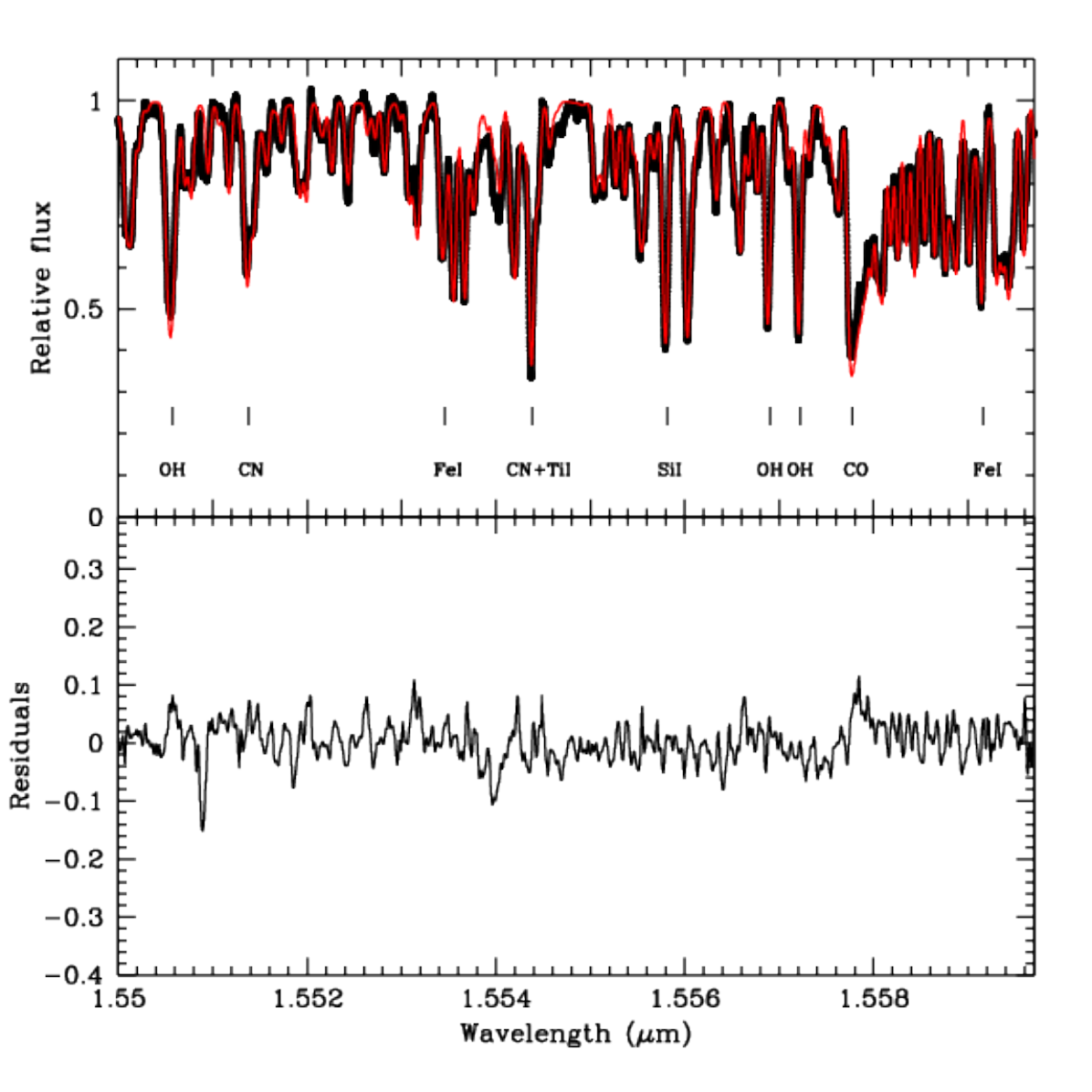}
   \caption{Top panel: Comparison between the  observed (black line) and synthetic spectra (red line, best fit) for the star V715 Lyr belonging to NGC 6791 in the spectral range $\sim 1.55$ $\mu$m. Some of
the atomic and molecular lines used in the chemical analysis are indicated. In the figure, for $\lambda\lambda > 1.558$ $\mu$m,
most of the absorptions are due CO molecular lines. Bottom panel: Residuals of the fit.} 
   \label{fig:fig1}
\end{figure}

\begin{figure}
\centering
\includegraphics[width= 8 cm]{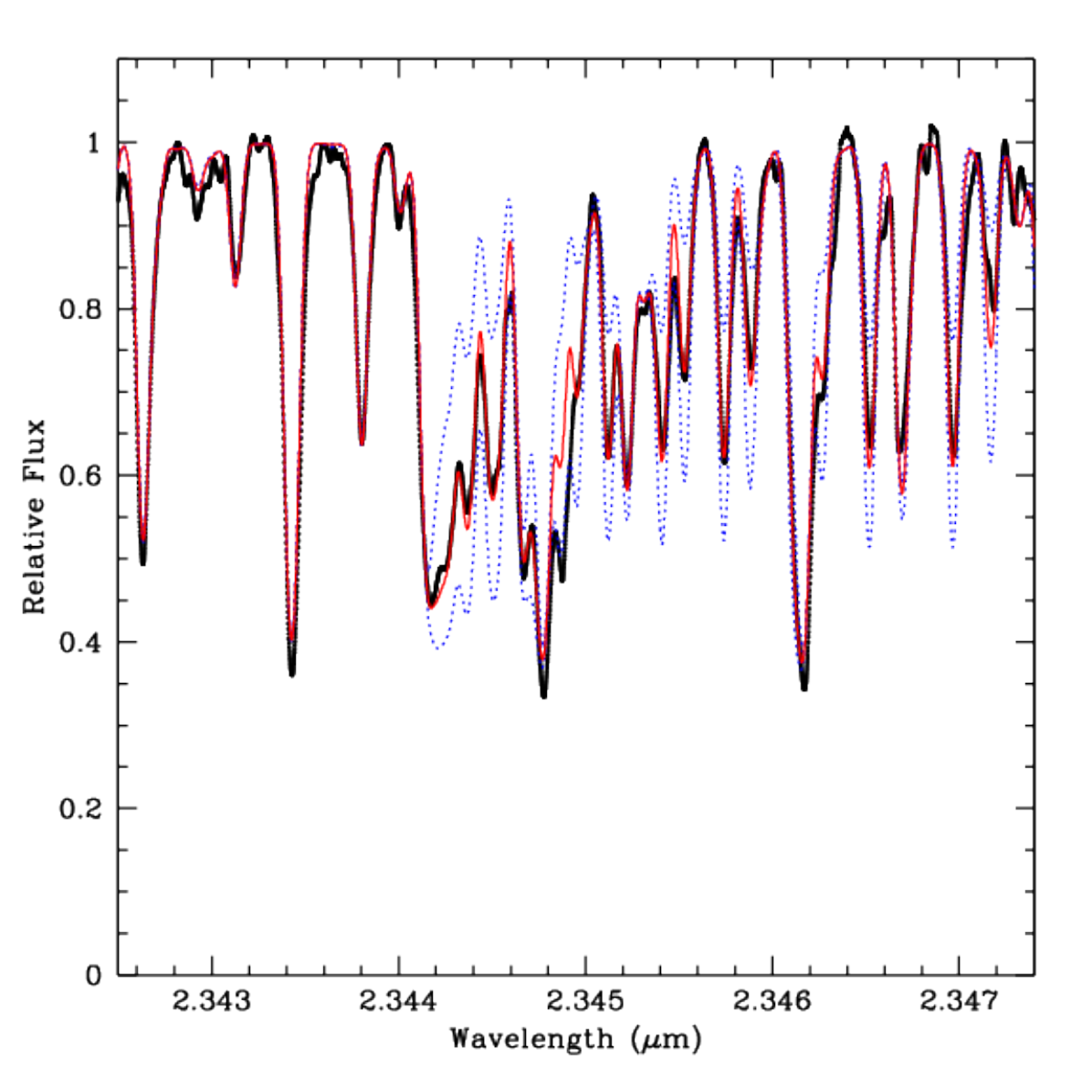}
\caption{Comparison between the  observed (black line) and several synthetic spectra (colour lines) for the RGB star 494 (NGC 7789) in the region of the $^{13}$C$^{16}$O molecular band from where the carbon isotopic ratios were
derived. Blue-dotted lines are theoretical spectral
with $^{12}$C/$^{13}$C$=100$ and 10, respectively. Red continuous line (best fit) is computed with 
$^{12}$C/$^{13}$C$=21$.}
\label{fig:fig2}
\end{figure}

\begin{figure}
\centering
\includegraphics[height=8 cm,angle=00]{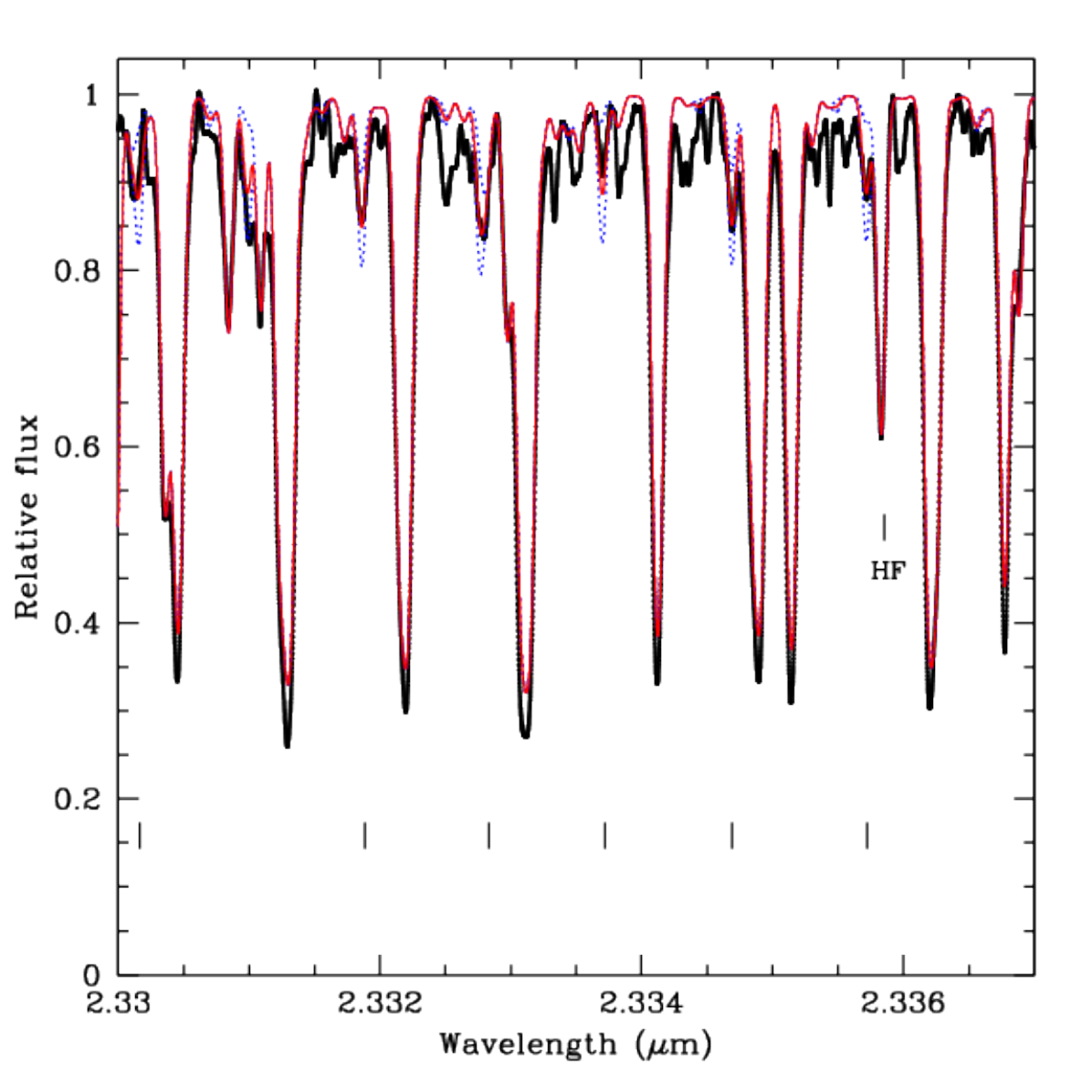}
\caption{Comparison between the  observed (black line) and several synthetic spectra (colour lines) for the RGB star KIC 5199859 (NGC 6819) in the spectral region where most of the $^{12}$C$^{17}$O lines used are found
(marked with a line). Also the R9 HF molecular line is indicated. The strong spectral features are $^{12}$C$^{16}$O lines. Synthetic spectra (blue dotted lines) calculated for $^{16}$O/$^{17}$O$=2000$, 400 and 800 (red continuous line), respectively.}
\label{fig:fig3}
\end{figure}

Next, we derive the $^{12}$C/$^{13}$C ratio in the stars using the absorption of the $^{13}$CO band at $\lambda\lambda\sim 2.345$ $\mu$m (see Fig. 6). This band contains many $^{13}$C$^{16}$O lines that are very sensitive to variations in the abundance of the $^{13}$C isotope. Some features of the $^{13}$CO molecule at $\lambda\lambda\sim 1.65-1.67$ $\mu$m provided a secondary estimate of the isotopic carbon ratio. For this, we also used the atomic and molecular line list in the APOGEE DR17 project \citep{smi21}. Good agreement was obtained within the uncertainties (see below) between the $^{12}$C/$^{13}$C ratio obtained from both spectral ranges. Finally, we used again the $\chi^ 2$ minimisation technique to determine the oxygen ratios that provide the best fit to each C$^{17}$O line. An example can be seen in Figure 7. The oxygen isotopic ratios derived from individual features were then combined to obtain an average (see Table 3). The number of spectral features used in each star is also shown in Table 3, together with the average value obtained for $^{16}$O/$^{17}$O, and the maximum and minimum values derived for each star. The latter is a direct estimate of the dispersion in the derived isotopic ratio. 
In general, this dispersion is quite low except for the stars 3012 and V715 Lyr in NGC 6791. Indeed, the spectra of these two stars are the most crowded among the
stellar samples studied probably because of their relatively low
 S/N ratio of their spectra (see below). 
 These stars have the highest oxygen isotopic ratios found in the stellar sample, which is consistent with NGC 6791 being the oldest open cluster studied (see discussion below).

\begin{table*}
\tiny
\caption{Stellar parameters, abundances, isotopic ratios and luminosities derived for the RGB stars in the OCs sample.}                  
\centering          
\begin{tabular}{l c c c r c c r c r c}     
\hline\hline       
                      
Cluster& Star & T$_{\rm {eff}}$ & log g & [M/H] & $\xi$ & C/N/O & [O/Fe]& $^{12}$C/$^{13}$C & $^{16}$O/$^{17}$O &  M$_{\rm{bol}}$\\ 
&      &  (K)    & (dex) &  (dex)   &  (km/s)& & (dex)&  &   & (mag)\\ 
\hline 
 & & & & & & & & & & \\
NGC 6791 &  V715 Lyr  & 3510 & 0.82 & 0.25&1.70 &8.75/8.36/9.06 & $0.08$& 16 &1800$^{+700}_{-700}$ (4)&  $-2.93$\\

R$_{gc}=7.65$ kpc        & 3012&3531 &0.83 &0.30 &1.75 &8.78/8.37/9.08 & $0.10$ &18 & 1421$^{+250}_{-150}$ (4)& $-2.94$\\
& & & & & & & & \\
NGC 6819& 960 &3400 &0.71 & $-0.02$& 1.90 &8.35/7.96/8.65&0.00&  24& 655$^{+40}_{-35}$ (6)& $-3.21$\\
R$_{gc}=7.80$ kpc   & KIC 5199859& 3590& 0.75&$-0.02$ &1.90  &8.22/7.94/8.64 &$-0.01$&26 &720$^{+80}_{-80}$ (7)& $-3.48$\\
& & & & & & & & &\\
NGC 7044& UCAC4 663-096810& 3750&1.20 &0.10 & 1.90&8.44/8.09/8.69 &$-0.08$&20 & 532$^{+50}_{-60}$ (4) & $-2.84$\\
R$_{gc}=8.44$ kpc&UCAC4 663-096830 & 3680& 1.30 &0.06 &1.80  &8.35/8.10/8.60&$-0.13$& 21& 485$^{+20}_{-20}$ (6)& $-3.06$\\
& UCAC4 663-096943& 3800& 1.35 & 0.05&1.70 &8.32/8.15/8.69&$-0.03$&25 & 480$^{+80}_{-80}$ (4)& $-1.95$\\
& UCAC4 663-096588& 3877& 1.30 & 0.08&1.70 &8.35/8.15/8.67&$-0.08$& 20& 500$^{+10}_{-5}$ (6) & $-2.40$\\
                 & & & & & & & & & \\
NGC 7789& 415& 3850 &0.50 &$-0.03$ &1.80 & 8.30/8.09/8.64&0.00& 22& 537$^{+130}_{-137}$ (5)&$-2.85$\\
R$_{gc}=9.14$ kpc   & 494 &3802 &1.20 &0.02 & 1.90&8.35/8.00/8.66 &$-0.02$& 21 & 473$^{+80}_{-70}$ (5) & $-2.70$\\
& 72&3930 &1.43 & 0.00&1.85 &8.28/8.04/8.65 &$-0.02$& 23& 487$^{+70}_{-50}$ (4)& $-2.19$\\
\hline\hline                  
\end{tabular} 

\tablefoot{Galactocentric distances are from \cite{cav24} based on Gaia DR3 astrometry. The names of the stars are taken from the SIMBAD database (see text).}
\label{tab:ocdata}
\end{table*} 

The derived oxygen ratios
are affected by a number of uncertainty
sources. In particular, uncertainty in gravity and carbon abundance
has a significant impact on the $^{16}$O/$^{17}$O derived. For example, for stellar parameters typical in the stars studied (T$_{\rm{eff}}\sim 3700$ K, log g$\sim 1.0$, 
[M/H]$\sim 0.0$) an uncertainty $\Delta$[C/H]$= \pm0.1$ dex introduces an error in the oxygen ratio of $\mp 120$. Similarly, a typical variation of $\pm 0.2$ dex in gravity introduces uncertainty in the oxygen isotopic ratio of $\mp 200$. Uncertainties in the effective temperature ($\pm 100$ K) and metallicity ($\pm 0.1$ dex ) have a much lower impact ($\mp50$ and $\pm 65$, respectively). Typical uncertainties in nitrogen and oxygen abundances as well as in microturbulence  introduce a negligible error in the determination of the $^{16}$O/$^{17}$O ratio. Variations in the $^{12}$C$/^{13}$C and $^{16}$O/$^{18}$O ratios do not affect the derivation of $^{16}$O/$^{17}$O from the CO lines used. A similar analysis can be performed with respect to the total uncertainty in the derivation of the carbon isotopic ratio. In this case, the main error sources are the T$_{\rm{eff}}$ and the carbon abundance. Changes in $\pm 100$ K or $\pm 0.1$ dex in these parameters are translated into an uncertainty of $\pm 7$ in the $^{12}$C/$^{13}$C ratio; while
changes by $\pm 0.2$ dex in gravity and $\pm 0.1$ kms$^{-1}$ in microturbulence mean variations by $\pm 6$ and $\mp 3$ in the carbon isotopic ratio, respectively. Typical uncertainties in metallicity, oxygen and carbon abundances have almost no effect in the $^{12}$C$/^{13}$C derived. However, the estimate of the total
error in these ratios is more complex than the simple quadratic addition of all
of these errors, because most of the stellar parameters are coupled
to at least one of the others. For example, a change within the
uncertainty quoted  in the adopted T$_{\rm{eff}}$ would change the oxygen
ratios derived by a certain amount, but the resulting fit to
the global spectrum might not be acceptable. Therefore, such a
variation in T$_{\rm{eff}}$ should be discarded. In other words,
the co-variance terms in the estimate of the total error have to be
taken into account. In our case, and given the inter-dependence
between the stellar parameters, this is rather complex to estimate.
To take this into account, a procedure designed by \cite{cay04} was followed. Briefly, for a typical star in the sample, a
synthetic spectrum is computed by modifying a stellar parameter
by an amount according to its assumed uncertainty. Then, all the
other stellar parameters are changed within their uncertainties
until a good fit to the global spectrum is found. The difference
between the oxygen ratios derived using the first (best) adopted
set of stellar parameters and those derived with the new set of
’good’ stellar parameters may give an estimate of the real uncertainty.
The total error would then be that of the largest difference
in the oxygen (carbon) ratios obtained with the different combinations of
the stellar parameters. We have performed this analysis for several
stars in the sample. The error estimated in this way, added
quadratically with that due to the uncertainty in the continuum
position ($\sim 1 \%$) and the dispersion in the isotope ratios when
derived from different lines (see Table 3), would give an estimate
of the total error. This ranges from 140 to 200 and 4 to
6 for the isotopic ratios $^{16}$O/$^{17}$O and $^{12}$C/$^{13}$C, respectively (the larger
the derived isotopic ratio, the greater the total error).

\subsection{Luminosity}\label{sec:luminosity}
The luminosities of the stars were estimated from photometry and astrometry measurements by Gaia DR3 \citep{GaiaDR3} following the method described in \cite{delp26}, except for the stars V715 Lyr and 3012 belonging to NGC 6791, and  star 960 in NGC 6819. The stellar parameters of these three stars are out-side the applicability range of the \cite{delp26} method. For these objects, we instead used de calibration of the bolometric correction of the $\rm{K}$ magnitude with the colour index $\rm{(J-K)}$ by \cite{ker10} for O-rich giants. We adopted the same extinction value $A_V$ for the stars in a given open cluster according to \cite{cav24}. The main source of uncertainty in the derivation of the luminosity is the distance taken from \cite{bai21}. The total uncertainty in M$_{\rm{bol}}$
ranges from $\pm 0.15$ to $\pm 0.20$ mag, depending mainly on the error in the parallax and the adopted extinction. In fact, leaving aside the differential
extinction issue in a given OC, there are still large discrepancies in the estimation of OC extinction as a whole
within the literature \citep[see e.g.][]{cant20,hunt24,cav24}. For stars with bolometric corrections from \cite{ker10}, the total uncertainty in the bolometric magnitude may reach $\pm 0.25$ mag.

\section{Results and discussion}\label{sect:4}

In this section, we present the derived carbon and oxygen isotopic ratios for all the stars in our sample. Only a few previous studies have reported similar determinations; when available, we compare the values obtained in the literature with our results. Next, with the method described in Sect. \ref{sect:2}, we estimate the masses and ages of the stars.

Except for the two stars in NGC 6791, the measured \cratio~ ratios are between $20$ and $26$, values in very good agreement with the expectation for post-FDU stars of near-solar metallicity and mass in the range $1\text{--}2$ M$_\odot$. Although most of these stars are brighter than the RGB bump, they do not show the typical low \cratio~ expected when an extra-mixing process is at work (see Sect. \ref{sect:2}). All these stars show C depletion and N enhancement, while $[\mathrm{(C+N+O)/Fe}] \approx 0$, within a few hundredths of a dex. This is further evidence that these stars underwent the FDU, when the ashes of the CN cycle are moved to the stellar surface. The \cratio~ derived for the two RGB stars in NGC 6791 (namely, V715 Lyr and star 3012) is systematically lower, which is consistent with a moderate extra-mixing process reaching temperatures of $T_{\mathrm{max}} \sim 15$~MK. Note that this cluster is the oldest in our sample, with a metallicity larger than twice the solar value.

\begin{table}
\tiny
\caption{Initial mass and age for RGB stars in open clusters with nearly solar composition.}             
\centering          
\begin{tabular} {l c c c }     
\hline\hline    
                      
Cluster & Star& Mass (M$_\odot$) & Age (Ga)\\
\hline 
NGC 6819& 960   & $1.641^{+0.098}_{-0.071}$  & $2.23^{+0.35}_{-0.41}$ \\
         & KIC 5199859 & $1.612^{+0.082}_{-0.072}$& $2.37^{+0.39}_{-0.37}$ \\
         &  &  & \\
NGC 7044& UCAC4 663-096810   & $1.720^{+0.157}_{-0.091}$ & $1.90^{+0.40}_{-0.50}$  \\
        &  UCAC4 663-096830      & $1.756^{+0.189}_{-0.104}$ & $1.76^{+0.42}_{-0.54}$ \\
        & UCAC4 663-096943   & $1.760^{+0.192}_{-0.105}$  &$1.75^{+0.43}_{-0.54}$ \\
        & UCAC4 663-096588      & $1.74^{+0.180}_{-0.100}$  &$1.81^{+0.42}_{-0.53}$  \\
         & & & \\
NGC 7789& 415 & $1.717^{+0.153}_{-0.090}$ &$1.91^{+0.39}_{-0.50}$   \\
        & 494&$1.766^{+0.195}_{-0.107}$ & $1.72^{+0.43}_{-0.54}$\\ 
        &  72&$1.754^{+0.188}_{-0.103}$ &$1.77^{+0.42}_{-0.53}$ \\
\hline\hline                  
\end{tabular} 

\label{tab:other_clusters}      

\end{table} 

\subsection{NGC 7789}

We analysed the spectra of three RGB stars in this cluster. The resulting \oratio~ values range from 473 (for star 494) to 537 (for star 415). As discussed in Section \ref{sec:chemical}, we assume an uncertainty of $\pm150$ in the measured oxygen isotopic ratios. For two stars (494 and 72), the \oratio~ values are very similar. Note that these two stars are also those for which we obtained the smallest dispersion across five different lines. 
For 415 \citep{leb15} report a smaller value of $^{16}$O/$^{17}$O$=250$. Although both values agree within the respective error bar, the difference may impact our estimation of the stellar mass. The major difference between the two analyses is in the gravity adopted for this star, namely log g$=0.5$ here versus 1.2 in \cite{leb15}. The value we obtain for the carbon isotopic ratio in 415 ($\text{\cratio}=22\pm5$) agrees well with the one reported by \cite{leb15} (19$\pm8$), but disagrees with \cite{tau05} ($9\pm3$). The origin of this discrepancy is unclear. It should be noted that \citet{tau05} used weak CN lines in the $\lambda\lambda 8000$ \AA~region, which unfortunately lie outside our spectral range.

According to the measured [Fe/H] and C, N, and O abundances, we adopt a solar composition for these stars. The masses estimated from the measured oxygen isotopic ratio are reported in Table \ref{tab:other_clusters}. By combining the three independent measurements, a weighted average of the resulting masses is $1.741^{+0.101}_{-0.057}$ M$_\odot$, corresponding to an age of $1.81^{+0.24}_{-0.30}$ Ga for NGC 7789.   

\subsection{NGC 7044} 

Our sample includes IR spectra of four RGB stars in NGC 7044. We derive \oratio~ values ranging from 480 to 532 and \cratio{}s between 20 and 25. To our knowledge, no previous studies have reported carbon or oxygen isotopic ratios for this cluster. Overall, the dispersion among the \oratio\ values obtained from different CO lines is small, which confirms that the stellar parameters were properly selected. Nevertheless, we assume also for this cluster an uncertainty of $\pm150$ for the measured oxygen isotopic ratios.
The spectroscopic properties of this cluster, including the carbon and oxygen isotopic ratios, closely resemble those of NGC 7789. The resulting initial masses for each of the four RGB stars are listed in Table \ref{tab:OCage}. The corresponding weighted average mass is $1.742^{+0.089}_{-0.050}$ M$_\odot$, while for the age we obtain $1.81^{+0.21}_{-0.26}$ Ga. Notably, the errors that we find for these average values are substantially larger than the dispersion among the values obtained for the four stars, an occurrence likely due to an overestimation of the \oratio~ errors.  

\subsection{NGC 6819}

NGC 6819 is an excellent benchmark for testing our method. This cluster, generally considered slightly older than NGC 7789 and NGC 7044, was a target of the Kepler mission, whose asteroseismic data enabled accurate mass estimates for several RGB stars \citep{bas11,miglio2012}. Furthermore, this cluster contains several detached eclipsing binaries that host upper main-sequence stars \citep{bre16}.

In the TNG archive, we found infrared spectra of two RGB stars that are members of NGC 6819. The carbon and oxygen isotopic ratios were derived using 6 and 7 CO spectral lines, respectively. In this case as well, the measured \cratio~ values (24 for 960 and 26 for KIC 5199859) align with expectations for post-FDU stars. The oxygen isotopic ratios ($655\pm150$ and $720\pm150$), are $\sim 30\%$ higher than those derived for the RGB stars in NGC 7789 and NGC 7044. This result confirms the older age of NGC 6819. From the measure \oratio~ values, we derived the initial masses of the two RGB stars, yielding an average mass of  $1.625^{+0.063}_{-0.051}$ M$_\odot$. The current mass estimate for RGB stars in NGC 6819, as obtained from Kepler data, is $1.61 \pm 0.04$ M$_\odot$ \citep{miglio2012}, which is in excellent agreement with our result.  
On the other hand, according to \citet{bre16}, the primary component of the WOCS 23009 system (a detached eclipsing binary) has a mass of $\approx1.54$ M$_\odot$. Given that this star belongs to the upper main sequence, its estimated mass is consistent with the initial mass we derived for present-day RGB stars.
Finally, our estimate for the NGC 6819 age is $2.31^{+0.26}_{-0.28}$~Ga, also in agreement with the estimate from astererosismology \citep[2.0-2.4 Ga,][]{bas11} and eclipsing binaries \citep[$2.38\pm0.05(\rm{ran})\pm0.22(\rm{sys})$,][]{bre16}.

\begin{table}
\tiny
\caption{Initial mass and age for RGB stars in NGC 6791 (from the measured \oratio~ ratios). }             
\label{tab:6791}      
\centering          
\begin{tabular}{l c c c c c}     
\hline\hline       
                      
Star & [Fe/H] & [O/Fe]& Y  &  Mass (M$_\odot)$ & Age (Ga) \\ 
\hline 
V715 Lyr    &    0.3    &    0.0   &   0.275    & $1.36^{+0.10}_{-0.10}$  & $6.13^{+1.81}_{-1.33}$  \\
V715 Lyr    &    0.3    &    0.1   &   0.275    &  $1.39^{+0.09}_{-0.07}$  &  $5.89^{+1.06}_{-1.21}$  \\   
V715 Lyr    &    0.3    &    0.0   &   0.320     &  $1.28^{+0.10}_{-0.09}$  &  $5.44^{+1.65}_{-1.29}$  \\   
3012    &    0.3    &    0.0   &   0.275     &  $1.46^{+0.13}_{-0.10}$  &  $ 4.87^{+1.32}_{-1.33}$     \\   
3012    &    0.3    &    0.0   &   0.320     &  $1.38^{+0.13}_{-0.10}$  &  $ 4.22^{+1.29}_{-1.16}$     \\   
\hline \hline                
\end{tabular} 
\end{table} 

\subsection{NGC 6791}\label{sec:6791}

The \cratio~ values we obtain for the two RGB stars belonging to NGC 6791 are the lowest among those measured in our stellar sample. Specifically, we find $16\pm5$ and $18\pm5$, for V715 Lyr and 3012, respectively. Substantially lower values, $\sim9$, were found by \citep{szi18}, in spite of negligible differences in the adopted stellar parameters.
Likely, the origin of this discrepancy lies in the use of a different grid of model atmospheres: MARCS models in the present study vs ATLAS models \citep{kur79} in the previous one. Note that MARCS models are considered more appropriate for the study of cool stars \citep{gus08}. This difference may also explain the discrepancy in the derived C, N, and O abundances. On the other hand, a better agreement is found with the carbon ratio reported by \cite{mcc23} for V715 Lyr ($\text{\cratio}=13\pm\text{3}$). Note that the luminosity of both stars is substantially higher than that of the RGB bump, so it is possible that the activation of an extra-mixing process has reduced the carbon isotopic ratio below the expected post-FDU values. In any case, the extra-mixing required to reduce the carbon isotopic ratio down to the measured values is not extreme enough to imply a change in the oxygen isotopic ratio. 

Indeed, V715 Lyr and the star 3012 show the highest oxygen isotopic ratios among the RGB stars in our sample, which is consistent with NGC 6791 being the oldest OC studied here (see Tab. \ref{tab:OCage}). In particular, we find $1800$ and $1421$, for V715 Lyr and star 3012, respectively. No other studies report the oxygen isotopic ratio for RGB stars in NGC 6791. For these objects we find the greatest dispersion in the  oxygen isotopic ratio derived from different lines. This large dispersion might be a consequence of the particularly low S/N ratio of their spectra. Furthermore, the large difference between the \oratio ~ values found may also indicate a clear issue with these measurements. It may be due to:  a) inaccurate stellar parameters, b) inaccurate oscillator strengths, and/or c) the contamination with spectral blends.

Concerning the first hypothesis, the stellar parameters used here for these stars are quite similar to those in similar studies \citep{abd22,jon20,tin19,car19,haw16,gar21,pru11}. Nevertheless, we analysed these two stars using the atmospheric parameters suggested by the authors mentioned above, and in no case did we find a better fit to the global spectrum (according to the $\chi^2$ test) than the one obtained using the stellar parameters listed in Table 3. Concerning the second hypothesis, unfortunately we could not use a benchmark star to test the reliability of the spectroscopic parameters adopted for the C$^{17}$O lines. For instance, the spectra of the well known RGB stars $\alpha$ Boo and $\alpha$ Tau show extremely weak C$^{17}$O lines in the $\lambda\lambda\sim 2.3$ $\mu$m region (e.g., \cite{abia2012}) and an accurate derivation of the oxygen isotopic ratios in them is impossible from these lines. 
Finally, we cannot exclude that this large dispersion may be due to blends with atomic or molecular lines whose contribution becomes dominant in some of the selected C$^ {17}$O spectral features when the \oratio~ ratio is intrinsically large ($\gtrsim 1300$ or so). In this case, the C$^{17}$O lines become very weak and less sensitive to variations in the $^{17}$O abundance. The impact of these blends seems to be less important in stars with lower \oratio~ ratios, as proved by the significantly smaller dispersion we found for the RGB stars in the younger clusters studied here. Note that the impact of these hypothetical blends may be more significant as the metallicity of the star increases; in fact, the two stars in NGC 6791 are the highest-metal-rich 
analysed in this study. If this is the case, we might underestimate the oxygen isotopic ratio derived from some C$^{17}$O lines in stars with very high $^{16}$O/$^{17}$O ratios.

Bearing in mind this potential issue, we now attempt to derive the masses of the two RGB stars in NGC 6791, by adopting a conservative $\pm400$ uncertainty in both  measured \oratio~ values. According to our spectroscopic analysis, the models adopted for this purpose have $\text{[Fe/H]}=+0.3$, $\text{[O/Fe]}=0$ or $\text{[O/Fe]}=+0.1$, and Y=0.275 (solar) or Y$=0.32$. Note that an He enhancement is expected, in principle, at super-solar metallicity. The initial masses and ages estimate we obtain are summarised in Table \ref{tab:6791}. For comparison, the average current mass of the RGB stars in this cluster, as estimated from asteroseismic data, is $1.18$ M$_\odot$ $[\pm 0.01 \rm{(ran)} \pm 0.04 \rm{(sys)}]$ \citep{bas11, miglio2012}. Even taking into account the mass lost during prior evolution ($<0.1$ M$_\odot$), our mass estimates are definitely higher than those derived from asteroseismology. Better agreement is found by assuming an increased initial helium abundance and a solar-scaled chemical composition. In this case, our estimate agrees between the $1\sigma$ and $2\sigma$ error with the average asteroseismic mass, for V715 Lyr and star 3012, respectively. In principle, the existence of some extra-mixing could reduce the \oratio~ and, as a consequence, alleviate the discrepancy, but in this case the \cratio~ ratio should be $<9$, in contrast to the measured values. As previously discussed, the uncertainty adopted in our analysis does not include potential systematic errors that could lead to a non-negligible underestimation of the \oratio~ ratio due to spectral blends. In the future, acquiring infrared spectra with a higher signal-to-noise ratio and (possibly) spectral resolution would help uncover blended $\text{C}^{17}\text{O}$ lines that might otherwise be mistaken for higher $^{17}\text{O}$ abundance. Alternatively, the derivation of the oxygen isotopic ratios from the $\sim 5$ $\mu$m spectral range is suggested \citep{abia2012}. 

\section{Summary and conclusions}\label{sect:conclusions}
\label{sec:conclusions}

In this paper, we have presented a novel tool to infer the initial mass and, consequently, the age of RGB stars, based on the measurement of the surface oxygen isotopic ratio \oratio~. Following the FDU, the stellar convective envelope extends into deep layers previously processed by the CNO-cycle. While the surface \cratio~drops rapidly to $\sim25$, showing a weak variation with the mass, the post-FDU \oratio~shows a remarkably strong and continuous dependence on the initial mass for stars in the $1.0 - 2.0\,M_\odot$ range (see Fig. 2). In addition, we have shown that, at variance with the \cratio, \oratio~ is substantially less sensitive to the operation of extra-mixing processes that eventually are active below the convective envelope in bright RGB stars. Indeed, it is modified only if the extra-mixing reaches temperatures as high as $\sim20\text{ MK}$, for which the \cratio~ ratio is already reduced to values as low as $<10$ (i.e., significantly below its post-FDU level). In any case, even in the presence of a mixing particularly deep, the degeneracy between the effects of the FDU and extra-mixing can be broken by a simultaneous fit of both carbon and oxygen isotopic ratios. This procedure is illustrated in Section \ref{sec:arcturus} for $\alpha\,\text{Boo}$ (Arcturus).

To demonstrate the feasibility and robustness of this method, we analysed archival high-resolution ($R = 50\,000$) near-infrared spectra acquired with the GIANO-B spectrograph at the TNG. The sample includes 11 RGB stars belonging to four Galactic open clusters covering a range of ages and metallicities: NGC 7789, NGC 7044, NGC 6819, and NGC 6791.

Our main findings are summarised as follows:
\begin{itemize}
    \item Validation in intermediate-age clusters (NGC 6819, NGC 7044, NGC 7789):
    For \object{NGC 6819}, the derived initial mass ($1.625^{+0.063}_{-0.071}\,M_\odot$) and age ($2.31^{+0.26}_{-0.28}\text{ Ga}$) are in excellent agreement with independent current mass determinations from \textit{Kepler} asteroseismology \citep{bas11,miglio2012} and eclipsing binary systems \citep{bre16}. For \object{NGC 7044} and \object{NGC 7789}, where asteroseismic data are not available, we derived weighted average ages of $1.81^{+0.21}_{-0.26}\text{ Ga}$ and $1.81^{+0.24}_{-0.30}\text{ Ga}$, respectively. Crucially, for all three clusters, our oxygen-isotope age determinations are systematically higher than those derived by recent machine learning and pattern recognition algorithms trained on synthetic colour-magnitude diagrams. This highlights how synthetic CMD fits can remain affected by residual degeneracies between distance, reddening, and metallicity.

    \item Limitations in old, metal-rich clusters (NGC 6791): 
    For the oldest cluster in our sample, \object{NGC 6791} ($>8 \text{ Ga}$), the archival TNG spectra suffered from a relatively low S/N ratio. For stars older than $\sim 3-4\text{ Ga}$, the surface abundance of $^{17}\mathrm{O}$ following the FDU is intrinsically very low, making faint $^{12}\mathrm{C}^{17}\mathrm{O}$ molecular lines used in this study highly susceptible to continuum placement errors and line blending with surrounding metallic and/or molecular species. Consequently, the derived \oratio~values carry large dispersions, preventing a reliable constraint on the initial stellar mass and age for this object.

    \item On the operation of extra-mixing processes. All nine RGB stars belonging to the three clusters NGC 6819, NGC 7044, and NGC 7789 present \cratio~ values in agreement with the post-FDU theoretical predictions. Furthermore, for the only two stars in this study (to our knowledge) with an estimate of the abundance of lithium (stars 72 and 494 in NGC 7789), the measured values, A(Li)=1.15 and 1.05, respectively \citep{nag23}, are perfectly consistent with the expected values after the FDU according to standard models of stellar evolution for initial masses of 1-3 M$_\odot$ and solar composition \citep[e.g.][]{Charbonnel2010,pal111}. In practice, we find no evidence for the activation of extra-mixing. This stands in contrast to their brightnesses  (see the last column of Table~\ref{tab:ocdata}), which are all higher than that of the RGB bump, expected at $M_{\text{bol}} \sim 0.25$ for a cluster of about 2 Ga. In contrast, the two RGB stars in NGC 6791 display lower carbon isotopic ratios ($10 < \text{\cratio} < 20$). These values would be compatible with the activation of a moderate extra-mixing process, which, however, is not deep enough to induce a modification of the oxygen isotopic ratio.  
\end{itemize}

Overall, this work successfully proves the viability of using the \oratio~ isotopic ratio as a powerful independent way to infer stellar ages. Future observational programmes could target larger samples of RGB stars per cluster to improve statistical confidence. Furthermore, acquiring high-S/N, higher-resolution IR spectra and, possibly, extending into the $5\,\mu\mathrm{m}$ spectral region where less blended and intense O-bearing molecular transitions reside, will be essential to accurately resolve the faint $^{17}\mathrm{O}$ features, also in the oldest stellar populations.

\begin{acknowledgements}
Part of this work was supported by the Spanish project PID2021-123110NB-I00 financed by MICIU/AEI /10.13039/501100011033 and
by FEDER, una manera de hacer Europa, UE. This work is based (in part) on data obtained from the TNG Archive  made with the Italian Telescopio Nazionale Galileo (TNG), operated on the island of La Palma by the Fundación Galileo Galilei of the INAF (Istituto Nazionale di Astrofisica) at the Spanish Observatorio del Roque de los Muchachos of the Instituto de Astrof\'\i sica de Canarias.
\end{acknowledgements}

  \bibliographystyle{aa} 
   \bibliography{oiso.bib} 

\begin{appendix}
\section{Impact of convective core overshooting  on the $^{16}\text{O}/^{17}\text{O}$--age relation}\label{A1}

Convective core overshooting during the main-sequence phase remains one of the largest sources of systematic uncertainty in stellar evolutionary models with a mass greater than $\sim 1.1$ M$_\odot$. Although turbulent plumes are universally accepted to penetrate beyond the formal Schwarzschild boundary ($\nabla_{\mathrm{rad}} = \nabla_{\mathrm{ad}}$), there is no general consensus on how to parametrise the extent, velocity distribution, and thermal structure of this overshooting zone. Note that this uncertainty impacts virtually all open cluster dating methods based on theoretical stellar models. In particular, it affects standard isochrone fitting as well as various asteroseismic age diagnostics, such as those that rely on RGB mass estimates or the main-sequence hydrogen consumption timescale.

Classical stellar models treated overshooting through a simple \emph{step-overshooting} scheme, extending the fully mixed convective core instantaneously by a fraction $\Lambda_{\mathrm{ov}}$ of the local pressure scale height
\begin{equation}
H_P = -\frac{\mathrm{d}r}{\mathrm{d}\ln P} = \frac{P}{\rho g}
\end{equation} 

This is the case, for example, of the PARSEC models \citep{bre12}, which are widely adopted for isochrone-fitting and other age-determination methods.
More modern evolutionary codes, following the results of numerical hydrodynamical simulations of convective boundaries (e.g. \cite{fre96}), treat overshooting as a time-dependent diffusive mixing process. In these models, the convective velocity and hence the diffusion coefficient decay exponentially beyond the Schwarzschild boundary. For example, the diffusion coefficient $D_{\mathrm{ov}}(z)$ as a function of the radial distance $z$ from the boundary is expressed as:
\begin{equation}
D_{\mathrm{ov}}(z) = D_0 \exp\left( -\frac{2z}{f_{\mathrm{ov}} H_P} \right),
\end{equation}
where $D_0$ is the diffusion coefficient evaluated within the convective core near the boundary (as derived from mixing-length theory) and $f_{\mathrm{ov}}$ is a free parameter that requires empirical calibration against observational constraints, such as cluster CMD turn-off morphology or double-lined eclipsing binary properties.

A further theoretical challenge lies in the treatment of the energy flux transported by convective motion within the overshooting layer. Conventional evolutionary models adopt the simple assumption that the turbulent convective heat flux is negligible in the overshooting region ($F_{\mathrm{conv}} \approx 0$). Consequently, the temperature gradient in this zone is assumed to be purely radiative ($\nabla = \nabla_{\mathrm{rad}}$). However, recent investigations \citep[e.g.][]{hig21, ahl22} have called this assumption into question, demonstrating that turbulent enthalpy transport in the overshoot region can remain non-negligible, thereby altering the thermal structure near the core edge.

Regardless of the adopted mixing scheme, a unique choice of the overshoot parameter ($\Lambda_{\mathrm{ov}}$ or $f_{\mathrm{ov}}$) across all stellar masses and compositions cannot satisfy all observational constraints. 
This issue is particularly evident for models in the transition regime between stars burning hydrogen through the p-p chain in a radiative core and those burning hydrogen primarily through the CNO cycle in a convective core. When a strict application of standard overshoot schemes is employed, a clear mismatch arises between the predicted and observed open cluster CMD morphology  
 \citep{apa93, pie04,Magic2010} and other constraints, e.g., from eclipsing binaries \citep{cla16}, and asteroseismic data of $\gamma$~Dor stars \citep{mom19}.
Indeed, for stars with masses $M \lesssim 2 \, M_\odot$, which are the primary focus of our work, the physical radius and mass of the convective core are inherently small during central hydrogen burning. Because $H_P$ diverges as $r \to 0$, a straightforward application of Eq.~(2) using the pressure scale height at the Schwarzschild boundary tends to artificially overestimate the volume of the overshooting zone. 
 For example, \cite{hig21} found that for intermediate-mass stars, $f_{\text{ov}}$ within $0.01 < f_{\text{ov}} < 0.017$ satisfies observational constraints. However, as stellar mass and convective core size decrease, lower $f_{\text{ov}}$ values should be adopted (e.g., $f_{\text{ov}} < 0.005$ for a $1.3$ M$_\odot$ star).
In practice, this limitation 
can be implemented via a mass-dependent 
$f_{\text{ov}}$ or, more physically, by constraining 
the overshooting region to the geometry of the 
convective core itself.

The convective core overshooting in stars with small convective cores currently relies on at least three free parameters: the mixing length parameter ($\alpha_{\mathrm{MLT}}$), which determines the thermal structure within the convective unstable zone as well as the convective velocity or diffusion coefficient near the core boundary; the overshooting efficiency parameter ($f_{\mathrm{ov}}$), which governs the decay scale of the mixing in the overshooting region; and a geometric scaling factor, introduced to limit the overshooting efficiency in case of a tiny core size. Interestingly, calibrating these three parameters against specific observational constraints represents a particularly challenging task.

Keeping in mind the limitations mentioned above, we nonetheless aimed to explore the potential impact that convective core overshooting would have on the method described in this work. To this end, we computed a new set of evolutionary tracks for stars with masses between $1.5$ and $1.8$  M$_\odot$ and solar chemical composition, explicitly incorporating the effects of core overshooting.
With the stellar evolution code used in this work, FuNS, it is possible to choose among three different time-dependent mixing schemes, namely diffusive mixing, advective mixing, and mixing proportional to a function of the ratio between the time step and the mixing timescale \citep[see][]{stra06}. For this test, we adopted an advective-type mixing scheme, incorporating an exponential decay of the convective velocity starting from the convective boundary\footnote{A test done assuming a diffusive scheme led to similar results.}.
Furthermore, we set $f_{\text{ov}} = 0.02$ for stars with $M \ge 2.5\,M_\odot$. For lower-mass stars, $f_{\text{ov}}$ is linearly rescaled down to zero at the radiative–convective core transition ($\sim 1.2\,M_\odot$) to prevent artificially large overshooting caused by the geometrical divergence of $H_P$.

The inclusion of convective core overshooting affects the resulting surface abundance ratios on the RGB through two main mechanisms:

\begin{enumerate}
    \item {Main-sequence lifetime:} The extra hydrogen mixed into the nuclear burning region prolongs the MS phase. As a result, for any given stellar mass, models with overshooting exhibit longer MS lifetimes (and larger post-MS ages) compared to non-overshooting tracks.
    
    \item {Internal $^{17}\mathrm{O}$ profile:} At central hydrogen exhaustion, models with core overshooting develop a slightly steeper radial gradient of $^{17}\mathrm{O}$ in the region partially processed by CNO burning. Although the maximum penetration depth (in mass coordinate) of the convective envelope during FDU is virtually identical between the two sets of models, the steeper inner profile of $^{17}\mathrm{O}$ leads to a higher post-FDU surface ratio of $^{16}\mathrm{O}/^{17}\mathrm{O}$ in overshooting models.
\end{enumerate}
   
\begin{figure}[htbp]
   \centering
   \includegraphics[width=\columnwidth]{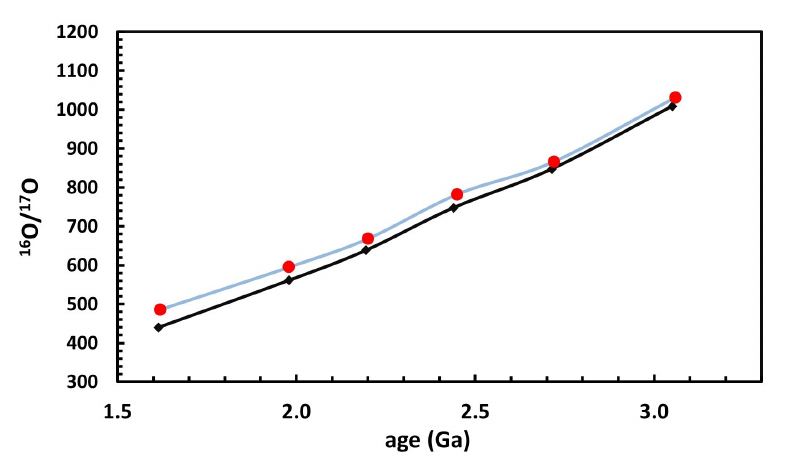}
   \caption{Theoretical post-FDU $^{16}\mathrm{O}/^{17}\mathrm{O}$ surface abundance ratio as a function of stellar age (in Ga). for models with $M = 1.5 - 1.8 \, M_{\odot}$. Black lines with diamond markers represent standard models without overshooting, while red lines with circular markers correspond to models incorporating convective core overshooting (see text for details).}
   \label{fig:ov_age}
\end{figure}

These combined effects are quantified in Fig. \ref{fig:ov_age}. 
For a given initial mass, models with overshooting exhibit systematically higher post-FDU $^{16}\text{O}/^{17}\text{O}$ values compared to standard models. Consequently, the mass associated with a given isotopic ratio value is larger. However, when converting to an age estimate, the longer main-sequence lifetime in overshooting models partially compensates for this variation. In general, for NGC 6819, adopting models with overshooting yields an age that is $\sim 3 \%$ younger, while for NGC 7044 and NGC 7789, the age reduction is 8\%. These variations are well within the uncertainties quoted and, therefore have no practical effect on our conclusions.

\section{The depth of the convective envelope at the first dredge up}\label{A2}

During the transition between the main sequence and the RGB, the increase in opacity in the outer layers, as caused by the expansion of the envelope, and the growing energy flux from the newly formed H-burning shell, drives the penetration of the outer convection, resulting in the modification of the CNO abundances at the stellar surface. When the H-burning shell approaches the inner boundary of the convective envelope, its entropy barrier halts the penetration of the instability and pushes the convection outward. As discussed in Sect. \ref{A1}, the formation and progression of the H-burning shell during the phase preceding the FDU are mainly affected by the chemical profile left behind by the convection of the core during the main sequence. \cite{Charbonnel2010} also investigated the effect of rotation on the FDU, finding that it induces a substantially greater lithium depletion, but only moderate reductions in both $^{12}\text{C}/^{13}\text{C}$ and $^{16}\text{O}/^{17}\text{O}$. 

In our calculations, we adopted a solar-calibrated mixing-length parameter. Our tests demonstrate that both mass and age estimations through the \oratio\ are only marginally affected by a variation of $\alpha_{\text{MLT}}$. Indeed, while the outer layers of the convective envelope develop a certain degree of super-adiabaticity, in the deeper layers, where the heat capacity is high, transport is virtually adiabatic. Specifically, in a $1.6\,M_\odot$ model with solar composition, when $\alpha_{MLT}$ is reduced from $1.873$ (the solar calibrated value) to $1.6$, the post-FDU $^{16}\text{O}/^{17}\text{O}$ ratio decreases (from $747$ to $722$), a difference compensated for by increasing the mass from $1.6$ to $1.611\,M_\odot$. Overall, the estimated ages are reduced by about 2\%.

Regarding the possibility of overshooting below the bottom of the convective envelope, it should be considered that the maximum penetration occurs when the H-burning shell is already very close to the convective boundary. Under these conditions, any potential overshooting is constrained due to steep entropy and mean molecular weight gradients. Obviously, parametric models adopting arbitrarily large overshooting can lead to substantial modifications in the post-FDU surface chemical composition. However, such models usually do not account for the thermal structure between the burning shell and the convective boundary, and should therefore be treated with extreme caution. As reported in Sect.~\ref{sec:conclusions}, the Li abundance measured in at least two stars of our sample is perfectly compatible with what is commonly found in models without envelope overshooting.

\section{The initial oxygen composition}\label{A3}

\begin{figure}[b]
\centering
\includegraphics[width=\columnwidth]{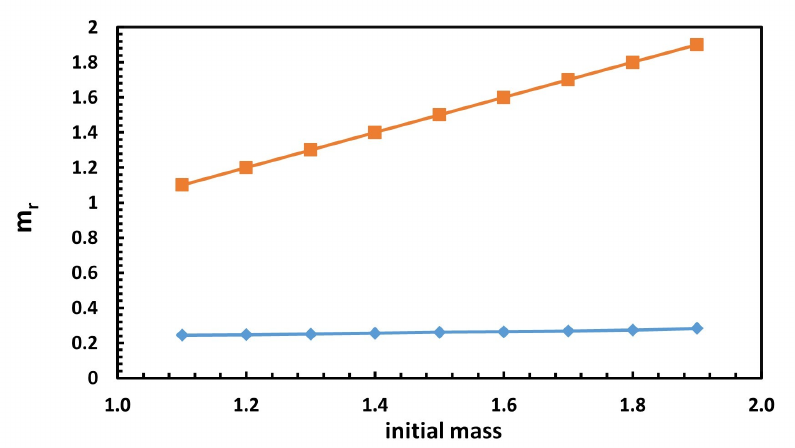} 
\caption{ depth reached by the convective envelope (blue line and diamonds) and total stellar mass at the time of the FDU (red line and squares), for models with solar chemical composition. The region between the two lines indicates the maximum mass extension of the convective envelope. Both axes are in solar masses (M$_\odot$).}
\label{FDU_depth}
\end{figure}

The FDU does not change the initial abundance of $^{16}\text{O}$, which can therefore be derived by combining the measurements of [O/Fe] and [Fe/H] (see Sect. \ref{sec:arcturus} and \ref{sect:4}). On the contrary, there are still no direct measurements that can constrain the initial abundance of $^{17}\text{O}$.
Therefore, to set the initial $^{17}\text{O}$ abundance, we adopted a standard protocol based on galactic chemical evolution  models and supported by observational constraints. According to standard GCE models \citep[e.g.,][]{Timmes1995,koba11,pra18}, $^{17}\text{O}$ behaves mainly as a secondary nucleosynthetic isotope. Consequently, its initial abundance is expected to scale with overall metallicity ($Z$ or $[\text{Fe/H}]$). As illustrated in Sect. \ref{sect:4}, the three youngest clusters in our sample show a practically solar chemical composition, specifically $\text{[Fe/H]}\sim 0$ and $\left[ \frac{\text{C}+\text{N}+\text{O}}{\text{Fe}}\right ] \sim 0$. Therefore, for these clusters we assumed a solar abundance of $^{17}\text{O}$. In contrast, NGC 6791 presents a super-solar metallicity, $\text{[Fe/H]}=0.3$, so the initial abundance of $^{17}\text{O}$ was rescaled to the highest iron content.
This secondary behaviour and metallicity scaling are generally supported by radio spectroscopy measures of oxygen isotopes in molecular gas across the Galactic disc \citep[see, e.g.,][]{Wouterloot2008,Zhang2020}.    

How would our predictions change in the case of deviations from this standard scenario? The post-FDU surface isotopic ratio is determined by mixing of material partially processed by H-burning in the deeper regions with the unburned outer envelope. Because the outer envelope is not nuclear-processed, its $^{17}\text{O}$ content remains fixed at its initial value. The surface abundance after FDU is therefore a direct outcome of this dilution process. In practice, the surface abundance after the FDU depends on the $^{17}\text{O}$ enhancement in the innermost region reached by surface convection, as well as on the ratio between the mass extent of this region and that of the unprocessed outer envelope. In principle, mass loss may affect the mass of the unprocessed material, but its effect should be negligible at the time of the FDU.    

Fig. \ref{FDU_depth} shows the maximum depth attained by the convective envelope and the total stellar mass at the time of the FDU, for models with solar composition.
Overall, the influence of the initial $^{17}\text{O}$ abundance on the post-FDU isotopic ratio depends on the stellar mass.
For example, in a $1.6$ M$_\odot$ model, a 10$\%$ increase in the initial $^{17}\text{O}$ mass fraction results in a relative decrease of $\sim -2.5\%$ in the post-FDU $^{16}\text{O}/^{17}\text{O}$ ratio. However, changing the initial mass results in variations of $-1.5\%$ and $-3.5\%$ for the 1.4 M$_\odot$ and 1.8 M$_\odot$ models, respectively.

\section{The simultaneous fit of the C and O isotopic ratios}\label{A4}

As illustrated in Sect.~\ref{sec:arcturus}, RGB stars brighter than the RGB bump that show $^{12}\text{C}/^{13}\text{C} \ll 20$ likely host an extra-mixing process operating below the convective envelope. This process can alter the \oratio\ left by the FDU. In this case,  a simultaneous fit of both C and O isotopic ratios provides the necessary constraints to infer the initial mass of the RGB star and, in turn, its age. In the presence of extra-mixing, besides the initial mass (here denoted as $x$), these two isotopic ratios also depend on the maximum temperature attained by this process (here $y$). Recall that extra-mixing sets in when the hydrogen-burning shell erases the steep molecular weight gradient left by the FDU.

In practice, we have followed an iterative procedure,  specifically, a multivariate Newton-Raphson method. In the following, $A(x,y)$ and $B(x,y)$ are the \cratio\ and \oratio, respectively. For instance, in the case of $\alpha\text{-Boo}$, the location of the star in the HR diagram  and the measured \cratio provide an initial guess $(x_0=1.2 \text{M}_\odot, y_0=20 MK)$. Then we solve the following matrix equation: 

\begin{equation}
    \begin{pmatrix} \Delta A \\ \Delta B \end{pmatrix} \approx \begin{pmatrix} \frac{\partial A}{\partial x} & \frac{\partial A}{\partial y} \\ \frac{\partial B}{\partial x} & \frac{\partial B}{\partial y} \end{pmatrix}_{(x_0,y_0)} \begin{pmatrix} \Delta x \\ \Delta y \end{pmatrix} = J \begin{pmatrix} \Delta x \\ \Delta y \end{pmatrix},
\end{equation}    
where $\Delta A = A_{\text{meas}} - A(x_0, y_0)$ and $\Delta B = B_{\text{meas}} - B(x_0, y_0)$.
The parameter updates are obtained by solving for the corrections:
\begin{equation}
    \begin{pmatrix} \Delta x \\ \Delta y \end{pmatrix} = J^{-1} \begin{pmatrix} \Delta A \\ \Delta B \end{pmatrix}.
\end{equation}    
Since the system is non-linear, this process is iterative: update $(x_{k+1}, y_{k+1}) = (x_k + \Delta x, y_k + \Delta y)$ and re-evaluate $J$ until $(\Delta x, \Delta y)$ converges. At each iteration, the partial derivatives in the Jacobian matrix are computed numerically using finite differences. It requires three stellar trachs each iteration, even if the partial derivatives do not change much after the second.  We stop the iterations when both $\Delta A/A$ and $\Delta B/B$ became smaller than $10^{-2}$.
Due to the weak dependencies of the carbon isotopic ratio on the initial mass ($\frac{\partial A}{\partial x} \approx 0$), convergence is usually fast (typically 3 or 4 iterations).

To estimate the uncertainties in the derived parameters $x$ and $y$ ($\sigma_x$ and $\sigma_y$) from the measurement errors $\sigma_A$ and $\sigma_B$, we apply standard first-order error propagation. Assuming independent measurement errors, the parameter uncertainties are given by:

\begin{equation}
\sigma_x \approx \sqrt{ \left( \frac{\partial x}{\partial A} \sigma_A \right)^2 + \left( \frac{\partial x}{\partial B} \sigma_B \right)^2 },
\end{equation}

\begin{equation}
\sigma_y \approx \sqrt{ \left( \frac{\partial y}{\partial A} \sigma_A \right)^2 + \left( \frac{\partial y}{\partial B} \sigma_B \right)^2 }.
\end{equation}

The inverse partial derivatives are the elements of the inverted Jacobian matrix $J^{-1}$. By mapping the components of $J^{-1}$ to these respective inverse derivatives, we obtain the full explicit expressions for the parameter uncertainties:

\begin{equation}
\sigma_x \approx \frac{1}{|\det(J)|} \sqrt{ \left( \frac{\partial B}{\partial y} \cdot \sigma_A \right)^2 + \left( \frac{\partial A}{\partial y} \cdot \sigma_B \right)^2 },
\end{equation}

\begin{equation}
\sigma_y \approx \frac{1}{|\det(J)|} \sqrt{ \left( \frac{\partial B}{\partial x} \cdot \sigma_A \right)^2 + \left( \frac{\partial A}{\partial x} \cdot \sigma_B \right)^2 }.
\end{equation}

\end{appendix}

\end{document}